\documentclass[preprint2]{aastex}
\pdfoutput=1

\usepackage{amsmath}

\usepackage{natbib}
\usepackage{xspace}
\usepackage{dsfont}

\shorttitle{Universality of the Small-Scale Dynamo Mechanism}
\shortauthors{Moll et al.}

\newcommand{\Nabla}{\ensuremath{\vec{\nabla}}}
\newcommand{\ddrx}{\ensuremath{\mathrm{d}^3x}}
\newcommand{\dek}{\ensuremath{\mathrm{d}k}}

\newcommand{\abs}[1]{\ensuremath{\left|#1\right|}}
\newcommand{\mean}[1]{\ensuremath{\langle#1\rangle}}
\newcommand{\unit}[1]{\ensuremath{\,\mathrm{#1}}}

\newcommand{\Qrad}{\ensuremath{Q_\text{rad}}}
\newcommand{\taucorr}{\ensuremath{\tau_\text{corr}}}
\newcommand{\vrms}{\ensuremath{v_\text{rms}}}

\newcommand{\Pra}{\ensuremath{\mathit{P\hspace{-.1em}r}}}
\newcommand{\Pram}{\ensuremath{\Pra_\mathrm{m}}}
\newcommand{\Ray}{\ensuremath{\mathit{R\hspace{-.1em}a}}}
\newcommand{\Rey}{\ensuremath{\mathit{R\hspace{-.1em}e}}}
\newcommand{\Reym}{\ensuremath{\Rey_\mathrm{m}}}

\newcommand{\Alfven}{Alfv\'{e}n\xspace}
\newcommand{\Alfvenic}{Alfv\'{e}nic\xspace}
\newcommand{\MURaM}{\texttt{MURaM}\xspace}

\newcommand{\TT}{\ensuremath{\mathcal{T}}}
\newcommand{\TBB}{\ensuremath{\mathcal{T}_\mathrm{BB}}}
\newcommand{\TVV}{\ensuremath{\mathcal{T}_\mathrm{vv}}}
\newcommand{\TIV}{\ensuremath{\mathcal{T}_\mathrm{Iv}}}
\newcommand{\TIB}{\ensuremath{\mathcal{T}_\mathrm{IB}}}
\newcommand{\TRS}{\ensuremath{\mathcal{T}_\mathrm{rs}}}
\newcommand{\TSR}{\ensuremath{\mathcal{T}_\mathrm{sr}}}
\newcommand{\TVBT}{\ensuremath{\mathcal{T}_\mathrm{vBT}}}
\newcommand{\TVBP}{\ensuremath{\mathcal{T}_\mathrm{vBP}}}
\newcommand{\TBVT}{\ensuremath{\mathcal{T}_\mathrm{BvT}}}
\newcommand{\TBVP}{\ensuremath{\mathcal{T}_\mathrm{BvP}}}
\newcommand{\EK}{\ensuremath{E_\mathrm{K}}}
\newcommand{\EB}{\ensuremath{E_\mathrm{B}}}
\newcommand{\EV}{\ensuremath{E_\mathrm{v}}}
\newcommand{\lambv}{\ensuremath{\lambda_\mathrm{v}}}
\newcommand{\Emag}{\ensuremath{E_\text{mag}}}
\newcommand{\Ekin}{\ensuremath{E_\text{kin}}}

\renewcommand{\vec}[1]{\boldsymbol{#1}}

\begin{document}

\title{Universality of the Small-Scale Dynamo Mechanism}

\author{R. Moll\altaffilmark{1}, J. Pietarila Graham\altaffilmark{2}, 
        J. Pratt\altaffilmark{3}, R.H. Cameron\altaffilmark{1}, 
        W.-C. M{\"u}ller\altaffilmark{3}, 
        \& M. Sch{\"u}ssler\altaffilmark{1}}

\altaffiltext{1}{Max-Planck-Institut f\"ur Sonnensystemforschung,
   37191 Katlenburg-Lindau, Germany}

\altaffiltext{2}{Department of Applied Mathematics \& Statistics,
The Johns Hopkins University,
Baltimore, MD 21218, USA}

\altaffiltext{3} {Max-Planck-Institut f{\"u}r Plasmaphysik, Boltzmannstr. 2,
   85748 Garching, Germany}
 
\date{Accepted on May 2, 2011}

\begin{abstract}
We quantify possible differences between turbulent dynamo action in
the Sun and the dynamo action studied in idealized simulations.  For
this purpose we compare Fourier-space shell-to-shell energy transfer
rates of three incrementally more complex dynamo simulations: an
incompressible, periodic simulation driven by random flow, a
simulation of Boussinesq convection, and a simulation of fully
compressible convection that includes physics relevant to the
near-surface layers of the Sun.  For each of the simulations studied,
we find that the dynamo mechanism is universal in the kinematic regime
because energy is transferred from the turbulent flow to the magnetic
field from wavenumbers in the inertial range of the energy spectrum.
The addition of physical effects relevant to the solar near-surface
layers, including stratification, compressibility, partial ionization,
and radiative energy transport, does not appear to affect the nature
of the dynamo mechanism.  The role of inertial-range shear stresses in
magnetic field amplification is independent from outer-scale
circumstances, including forcing and stratification.  Although the
shell-to-shell energy transfer functions have similar properties to
those seen in mean-flow driven dynamos in each simulation studied, the
saturated states of these simulations are not universal because the
flow at the driving wavenumbers is a significant source of energy for
the magnetic field.
\end{abstract}

\keywords{magnetic fields -- magnetohydrodynamics (MHD) --
          Sun: photosphere -- turbulence -- dynamo}

\section{INTRODUCTION}
\label{sec:intro}

A turbulent 3-D flow in an electrically conducting fluid is
capable of generating a magnetic field by self-excited dynamo action
if the magnetic Reynolds number is sufficiently large.  In the absence
of rotation or large-scale shear flow, the energy of the generated
magnetic field resides predominantly at spatial scales significantly
smaller than the driving (integral) scale of the turbulence
\citep{Schekochihin:etal:2004}. Such dynamos are therefore usually
referred to as small-scale dynamos (SSDs). Turbulent plasmas are found
in nearly all astrophysical systems. Since the big spatial scales of
these systems imply large Reynolds numbers, SSDs are expected to
operate in most of them. This leads to a widespread magnetization of
astrophysical plasmas, the consequences of which can be profound
\citep[e.g.,][]{Schekochihin:Cowley:2006, Wang:Abel:2009,
Schleicher:etal:2010, King:Pringle:2010, Ryu:etal:2008}.

Owing to their small-scale nature, observational evidence for magnetic
fields generated by SSDs is difficult to obtain. Methods have been
suggested to obtain information about the turbulent magnetic field in
the interstellar and intracluster media
\citep[e.g.,][]{Waelkens:etal:2009}.  Evidence for such a field with
mixed polarity on subresolution scales in the solar photosphere is
obtained using the Hanle effect \citep{Trujillo:etal:2004,
Kleint:etal:2010} and, possibly, by considering the statistical
properties of the resolved fields measured through the Zeeman effect
\citep{Graham:etal:2009b,Graham:etal:2009a}.

Apart from these observational attempts, most research into SSDs has
followed theoretical \citep[e.g.,][]{Kazantsev:1968} or numerical
simulation approaches.  Simulations of SSD action were carried out for
a variety of physical settings, such as forced homogeneous
incompressible turbulence \citep{Meneguzzi:etal:1981}, Boussinesq
convection \citep{Cattaneo:1999}, anelastic convection
\citep{Brun:etal:2004}, up to fully compressible convection including
relevant physics in the solar near-surface layers, such as radiative
transport and partial ionization \citep{Voegler:Schuessler:2007,
Graham:etal:2010}. Given such a range of physical conditions, the
question arises of whether the mechanism of SSD action is universal or
is qualitatively different in the presence of additional physics,
e.g., as present in the near-surface layers of the Sun.  In this
paper, we therefore analyze three incrementally more complex
simulations of SSDs, namely, (1) a simulation of forced homogeneous
incompressible MHD turbulence, (2) a simulation of Boussinesq
(incompressible) convection, and (3) a simulation of compressible and
stratified solar near-surface convection. To compare the simulations,
we consider energy spectra and shell-to-shell energy transfer rates in
the kinematic growth phase and in the saturated state of the dynamo.
Shell-to-shell energy transfer analyses measure the exchanges of
kinetic and magnetic energies between different wavenumbers.
In the case of incompressible MHD, the method has been well studied
for dynamos as well as for decaying MHD flows
\citep{2001Dar,2005Debliquy,2005Mininni,2006Carati,2010Cho}.

\citet{Schekochihin2007} have raised the question whether there is an
essential physical difference between incompressible mean-flow-driven
dynamos \citep{Alexakis:etal:2005,2005Mininni} and those driven by
random flows with correlation times shorter than their own turnover
times. They suggested that, by measuring the shell-to-shell transfer
of a dynamo resulting from using the latter forcing, one should be
able to settle this question
in the following way: if inertial-range motions dominate the
amplification of the magnetic field, the dynamo is purely a property
of the inertial range and independent of any system-dependent
outer-scale circumstances. We call this condition universal in the
Kolmogorov sense.

\citet{Schekochihin2007} were concerned about the large role of
driving-scale motions in the dynamos studied by
\citet{Alexakis:etal:2005,2005Mininni} and postulated these features
were peculiar to the mean-flow driven case. However,
\citet{2006Carati} studied a non-mean-flow-driven dynamo for the
saturated state and found a strong nonlocal contribution from the
forcing-scale motions to all scales of the magnetic field (as seen for
mean-flow driving).

The paper is organized as follows. In Sect.~\ref{sec:methods}, we
describe the three simulation runs and give a brief account of the
shell-to-shell analysis.  This method requires a special treatment in
the case of the solar convection simulation, which is described in
Appendix A. The results of our analysis are reported in
Sect.~\ref{sec:results}.  We discuss them in
Sect.~\ref{sec:discussion} and give our conclusions in
Sect.~\ref{sec:conclusions}.

\section{METHODS}
\label{sec:methods}

\subsection{Homogeneous Turbulence (HoT)}

We use a pseudo-spectral FFT code \citep{Gomez:etal:2005a,
Gomez:etal:2005b, Mininni:etal:2010} to solve the incompressible MHD
equations in a periodic box with $L=2\pi$,
\begin{gather}
  \frac{\partial\vec{v}}{\partial t}+\vec{\omega}\times\vec{v}=
  -\Nabla p+\vec{j}\times\vec{b}+\nu\nabla^2\vec{v}+\vec{f},\notag\\
  \frac{\partial\vec{a}}{\partial t}=\vec{v}\times\vec{b}+\eta\nabla^2\vec{a},\notag\\
  \Nabla\cdot\vec{v}=\Nabla\cdot\vec{a}=0 .
\end{gather}
Here \(\vec{\omega}=\Nabla\times\vec{v}\) is the vorticity.  The
magnetic field is given in \Alfvenic units, \(\vec{b}
=\Nabla\times\vec{a} =\vec{B}/\sqrt{\rho\mu}\), with the vector
potential $\vec{a}$ calculated in the Coulomb gauge.  The forcing
$\vec{f}$ is an Ornstein--Uhlenbeck process:  the amplitudes of the
complex harmonic modes with \(2\le\abs{{\vec k}}\le3\) are evolved in
time according to
\begin{equation}
 C_{k_x,k_y,k_z,t+\delta t} = C_{k_x,k_y,k_z,t}
 \left(1-\frac{\delta t}{\taucorr}\right)
 +\sqrt{\frac{2A}{\taucorr}}\cdot\xi
\end{equation}
where $\taucorr$ is the correlation time (taken as unity), $A$ is
chosen such that $\vrms\approx1.1$, and $\xi$ is normally distributed
noise with variance $\delta t$.  The noise is randomized every time
step, $\delta t = 8\cdot10^{-4}$.  The values of the diffusivities are
$\nu=\eta=8.8\cdot10^{-4}$ in accordance with the resolution of the
simulation which has $512^3$ modes (without de-aliasing). The integral
scale of the motion is $L_0=1.9$, giving an integral-scale turnover
time of $\tau_L=L_0/\vrms\approx1.7$, and (kinetic and magnetic)
Reynolds numbers $\Rey=\Reym =\vrms L_0/\nu\approx2400$.  A
non-magnetic simulation is run until a turbulent statistical steady
state is reached after $\mathord{\approx}8\tau_L$; then, a seed field
of harmonic modes in the range $20\le\abs{\vec{k}}\le30$ is
introduced. After the kinematic phase of the magnetic field growth,
the run is continued (initially with lower resolution and Reynolds
numbers in order to save computing time) until a statistically
stationary, saturated state is reached.

\subsection{Boussinesq Convection (BC)}

In the Boussinesq approximation the fluid is treated as incompressible
except for the inclusion of buoyancy effects related to gravity.  The
nondimensionalized equations are
\begin{gather}
\label{vortreal}
\frac{\partial\vec{\omega}}{\partial t}
    - \Nabla \times (\vec{v} \times \vec{\omega} + \vec{j} \times \vec{b})
    = \nu \nabla^2 \vec{\omega} - \Nabla \theta \times \hat{\vec{g}}, \notag\\
\frac{\partial\vec{b}}{\partial t} 
    - \Nabla \times (\vec{v}\times\vec{b})
    =  \eta \nabla^2 \vec{b},\notag\\
\frac{\partial\theta}{\partial t} + (\vec{v}\cdot\Nabla) \theta
    = \kappa \nabla^2 \theta - (\vec{v} \cdot \Nabla) T_0,\notag\\ 
\Nabla \cdot \vec{v} = \Nabla \cdot \vec{b} = 0,\quad \Nabla \times \vec{b} = \vec{j}.
\end{gather}
The temperature $\theta$ represents fluctuations about an equilibrium
state with a mean vertical temperature gradient $\Nabla T_0$.  The
magnetic field is represented in \Alfvenic units with an \Alfven number
of one.

The BC simulation is a pseudospectral calculation performed at a
resolution of $512^3$ in a fully periodic box.  The amplitude of all
modes with $k_z=0$ is set to zero to prevent the exponential growth of
these modes \citep[elevator instability, see][]{Calzavarini:etal:2006}
in the vertically periodic box. This non-restrictive
``pseudo-Rayleigh--B\'enard'' boundary condition inhibits the formation
of boundary layers that appear with Rayleigh--B\'enard boundary
conditions in a vertically closed box.  

The BC simulation was carried out for a Prandtl number $\Pra=
\nu/\kappa=1$ and magnetic Prandtl number $\Pram= \nu/\eta=2$.  The
magnetic Reynolds number, defined in terms of the integral scale $L_0$
and $\vrms$ is $Re_m=\vrms L_0/\eta \approx 4000.$ The Rayleigh number
is determined using the characteristic length-scale of the vertical
temperature gradient and calculated to be $\Ray=1/\nu \kappa =
5.0\cdot 10^5$.  Defined in this way, the Rayleigh number is not
simply comparable to the Rayleigh number of a system with defined
boundaries, but nevertheless gives an indication of the balance
between buoyancy and dissipative forces in the simulation.

The initial state of the BC simulation consists of fully-developed
hydrodynamic convection.  Random fluctuations of magnetic field, small
compared to the kinetic energy of the system, are seeded into the
lowest 16 spectral modes in order to observe the onset of the
linear phase where magnetic-field energy grows due to turbulent dynamo
action. After nonlinear saturation the system enters an energetically
quasi-stationary state.

\subsection{Compressible Solar Convection}

We use results from a dynamo simulation of near-surface solar convection
carried out with the \MURaM code \citep{Voegler:etal:2005}. The simulation
differs from the more idealized simulations of MHD turbulence described
above by including physical processes that are relevant for solar
convection: compressibility, stratification, radiative energy transport,
and partial ionization.  The equations treated with the
\MURaM code, written in conservation form, are
\begin{gather}
\frac{\partial\varrho}{\partial t} + \Nabla\cdot(\varrho\vec{v}) = 0 \notag\\
\begin{split}
   \frac{\partial\varrho\vec{v}}{\partial t}
   + \Nabla\cdot \left[ \varrho\vec{v}\vec{v} 
   + \left(p+\frac{\abs{\vec{B}}^2}{2\mu}\right)
    \mathds{1} -\frac{\vec{B}\vec{B}}{\mu} \right] \\
    = \varrho \vec{g}+\Nabla\cdot\underline{\underline{\tau}} ,
\end{split} \notag\\
\frac{\partial \vec{B}}{\partial t} 
    +\Nabla\cdot\left(\vec{v}\vec{B}-\vec{B}\vec{v}\right)
    = -\Nabla\times(\eta\Nabla\times\vec{B}), \notag\\
\begin{split}
    \frac{\partial e}{\partial t}
    + \Nabla\cdot\left[ \vec{v}\left(e+p+\frac{|\vec{B}|^2}{2\mu}\right) 
    -\frac{1}{\mu}\vec{B}( \vec{v}\cdot\vec{B} ) \right] \\
    = \frac{1}{\mu} \Nabla \cdot (\vec{B}\times \eta\Nabla\times\vec{B} )
    + \Nabla\cdot(\vec{v}\cdot\underline{\underline{\tau}}) \\
    + \Nabla\cdot(K\Nabla T) + \varrho (\vec{g}\cdot\vec{v}) + \Qrad.
\end{split}
\end{gather}
where $\varrho$ is the density, $p$ is the gas pressure, and $\vec{g}$
is the gravitational acceleration.  \(\vec{v}\vec{v}\),
\(\vec{B}\vec{B}\), \(\vec{v}\vec{B}\) and \(\vec{B}\vec{v}\) are dyadic
products, and $\mathds{1}$ is the $3\times 3$ unit matrix. The viscous
stress tensor, $\underline{\underline{\tau}}$, is written for a
compressible medium with a viscosity coefficient containing a
shock-resolving and a hyperdiffusive part \citep[for details,
see][]{Voegler:etal:2005}.  The total energy density per volume, $e$, is
the sum of internal, kinetic and magnetic energy densities.  $T$ is the
temperature and $K$ the thermal conductivity. The source term $\Qrad$,
which accounts for radiative heating or cooling, is determined by
integrating the equation of radiative transfer over a number of
directions for each grid cell as described in \cite{Voegler:etal:2005}.
The system of equations is completed by the equation of state, which
describes the relations between the thermodynamical quantities of a
partially ionized fluid.

The numerical procedure uses centered, fourth-order explicit finite
differences for the spatial derivatives on a uniform Cartesian grid and
a fourth-order Runge--Kutta scheme for the time stepping. The boundary
conditions are periodic in the horizontal directions with a closed,
free-slip top boundary and an open lower boundary that permits the
free in- and outflow of fluid \citep[for details,
see][]{Voegler:2003,Voegler:etal:2005}.

Here we consider the results of the simulation that has previously
been presented as Run~C in \citet{Voegler:Schuessler:2007} and Run~2
in \citet{Graham:etal:2010}.  The computational box represents a
rectangular domain with an extent of \(4.86\unit{Mm}\) in the
horizontal directions and \(1.4 \unit{Mm}\) in the vertical direction,
covering the range from \(800\unit{km}\) below to \(600\unit{km}\)
above the optical solar surface\footnote{Unlike the other two
simulations presented in this paper, the simulation of solar
convection is not scale-invariant.  The equation of state and the
radiative transfer depend on explicit physical scales.}.  The
finite-difference scheme uses \(648\times648\times140\) grid cells,
corresponding to a grid spacing of \(7.5\times7.5\times10\)
kilometers.

The magnetic Reynolds number for this simulation is \(\Reym =\vrms
L_0/\eta\approx2000\).  Because the viscous stress tensor
$\underline{\underline{\tau}}$ is based on shock-resolving and
hyperdiffusivity, it is not simple to give explicit values of the
Reynolds or magnetic Prandtl numbers. An estimate of the magnetic
Prandtl number of between 1 and 2 was derived in
\cite{Graham:etal:2010} by considering the Taylor scales of the flow
and magnetic field. The Reynolds number $\Rey$ is then in the range of
1000 to 2000.

\subsection{Shell-to-Shell Transfer Analysis}
\label{sec:transan}

This method was first derived for incompressible Navier-Stokes
by \citet{1953Batchelor} and for incompressible MHD by
\citet{2001Dar}.  We follow here the exposition by
\citet{Alexakis:etal:2005} who previously applied the
analysis to kinematic and saturated dynamo states of homogeneous
turbulence driven by a mean flow.  We extend their study to random
forcing as well as to more physically realistic simulations by
deriving the compressible MHD shell-to-shell transfer functions.

The analysis starts with a decomposition of the velocity field and the
magnetic field according to \( \vec{a}(\vec{x}) = \sum_K
\vec{a}_K(\vec{x}) \), where \(\vec{a}_K\) is the part of $\vec{a}$
whose three-dimensional wave vector $\vec{k}$ in Fourier space lies in
the range \(K < \abs{\vec{k}} \le K+1\). The interval \((K,K+1]\) is
referred to as ``shell $K$''.  Logarithmic binning
$(\gamma^nK_0,\gamma^{n+1}K_0]$ for $\gamma>1$ and integer $n$ is
required to determine scale-to-scale energy transfers
\citep{Eyink2009}.  However, we are not seeking to answer questions of
scale locality of dynamo mechanisms (see, instead,
\citealt{2006Carati}).  Instead, we seek to determine whether the
mechanism seen in incompressible homogeneous dynamos is also at work
in our simulations.

For this purpose we use linear binning as opposed to the coarser
analysis resulting from logarithmic binning; logarithmic binning
can be recovered by summing over linear bins.
The spectral rates of change of the kinetic and
magnetic energies can then be written as
\begin{align}
  \frac{\partial}{\partial t}  E_\mathrm{v}(K) &= 
      \sum_Q \bigl[ \TVV + \TBVT + \TIV \bigr] + \mathcal{F}, \\
  \frac{\partial}{\partial t}  E_\mathrm{B}(K) &= 
      \sum_Q \bigl[ \TBB + \TVBT + \TIB \bigr],
\end{align}
where \(\TT=\TT(Q,K)\) are transfer functions:
\begin{align}
\label{eq:transvv}
    \TVV (Q,K) &= - \int \vec{v}_K (\vec{v} 
        \cdot \Nabla) \vec{v}_Q \, \ddrx,  \\
    \TBB (Q,K) &= - \int \vec{B}_K (\vec{v} 
        \cdot \Nabla) \vec{B}_Q \, \ddrx,  \\
    \TBVT(Q,K) &= \phantom{+} \int \vec{v}_K (\vec{B} 
         \cdot \Nabla) \vec{B}_Q \, \ddrx,  \\
    \TVBT(Q,K) &= \phantom{+} \int \vec{B}_K 
         (\vec{B} \cdot \Nabla) \vec{v}_Q \, \ddrx .
\label{eq:transvBT}
\end{align}
$\TIV$ and $\TIB$ represent the effects of dissipative heating, where
the index ``I'' stands for internal energy.
\(\mathcal{F}=\mathcal{F}(K)\) corresponds to the contribution of
external forces (e.g., gravity).  The first two indices of the $\TT$'s
denote the energy reservoirs involved in the transfer; $\TRS(Q,K)$ is the
rate of energy transferred from field $r$ in shell $Q$ to field $s$ in
shell $K$.  If $\TRS$ is positive, energy is received by $s$ from $r$
(transfer \(Q\rightarrow K\)). If it is negative, energy is lost by
$s$ to $r$ (transfer \(Q\leftarrow K\)).  The third index in $\TBVT$
and $\TVBT$ denotes the mediating force, in this case magnetic
tension.  Since Eqs.~\eqref{eq:transvv}--\eqref{eq:transvBT} express
energy transfers from field $r$ in shell $Q$ to field $s$ in shell
$K$, they satisfy the antisymmetry/conservation relations
\begin{equation}
    \TRS(Q,K) = - \TSR(K,Q).
\label{eq:anitcomp}
\end{equation}

We now introduce the shell-to-shell transfer functions for
compressible MHD.  In the compressible case, there is work done by
magnetic pressure, indicated by the index ``P''.  The work done by (or
against) magnetic pressure cannot be separated from energy transferred
between different wavenumbers inside the magnetic energy reservoir.
Thus, the above system of shell-to-shell transfer functions cannot be
used in the compressible case. As is consistent for antisymmetric
pairings, we write instead:
\begin{align}
  \frac{\partial}{\partial t}  \EK(K)
       &= \sum_Q \bigl[ \TVV + \TBVP + \TBVT + \TIV \bigr] + \mathcal{F},\\
  \frac{\partial}{\partial t}  \EB(K)
       &= \sum_Q \bigl[ \TVBP + \TVBT + \TIB \bigr] .
\end{align}
In addition to viscous dissipation, $\TIV$ now includes the internal
energy transferred to the kinetic energy reservoir by compression.
\citet{Graham:etal:2010} show that this new transfer, $\TVBP$, accounts
for 5\% of the magnetic energy generated in the \MURaM dynamo.

The transfer of magnetic energy in shell $Q$ to kinetic energy in shell
$K$ through the magnetic tension force is 
\begin{multline}
    \TBVT(Q,K) = \frac{1}{2\mu} \int \bigg[ \vec{v}_K \cdot (\vec{B} \cdot \Nabla) \vec{B}_Q \\
     + (\rho\vec{v})_K \cdot \left(\frac{\vec{B}}{\rho} \cdot \Nabla\right) \vec{B}_Q \bigg] \,\ddrx,
\end{multline}
where the integral is taken over the analysis domain.  The transfer of
kinetic energy in shell $Q$ to magnetic energy in shell $K$ through the
magnetic tension force is
\begin{multline}
    \TVBT(Q,K) = \frac{1}{2\mu} \int \vec{B}_K \cdot \bigg[ (\vec{B} \cdot \Nabla) \vec{v}_Q \\
                        +  (\vec{B} \cdot \Nabla) \frac{(\rho\vec{v})_Q}{\rho} \bigg] \,\ddrx .
\end{multline}
The transfers associated with magnetic pressure are
\begin{gather}
\begin{split}
\TBVP(Q,K) = - \frac{1}{2\mu} \int \vec{B} \cdot \bigg[ (\vec{v}_K \cdot \Nabla) \vec{B}_Q \\
                + \left( \frac{(\rho\vec{v})_K}{\rho} \cdot \Nabla \right) \vec{B}_Q \bigg] \,\ddrx,
\end{split} \\
\begin{split}
    \TVBP(Q,K) = - \frac{1}{2\mu} \int \vec{B}_K \cdot \bigg[ (\vec{v}_Q \cdot \Nabla) \vec{B} \\
                       + \left( \frac{(\rho\vec{v})_Q}{\rho} \cdot \Nabla \right) \vec{B} 
                        +  \vec{B} ( \Nabla \cdot \vec{v}_Q ) \\
                        + \vec{B} \left( \Nabla \cdot \frac{(\rho\vec{v})_Q}{\rho} \right) \bigg] \,\ddrx.
\end{split}
\end{gather}
The transfers due to different force groupings, here magnetic tension
(index T) and pressure (index P), separately satisfy the
conservative antisymmetry relation, Eq.~\eqref{eq:anitcomp}.
The application to non-periodic boundaries (\MURaM) necessitates a
windowing of the data, described in detail in the Appendix.

\section{RESULTS}
\label{sec:results}
\subsection{Field Morphology and Energy Spectra}

\begin{figure*}[t]
\centering
\begin{tabular}{c@{\hskip5pt}c@{\hskip5pt}c}
\includegraphics[height=.32\linewidth]{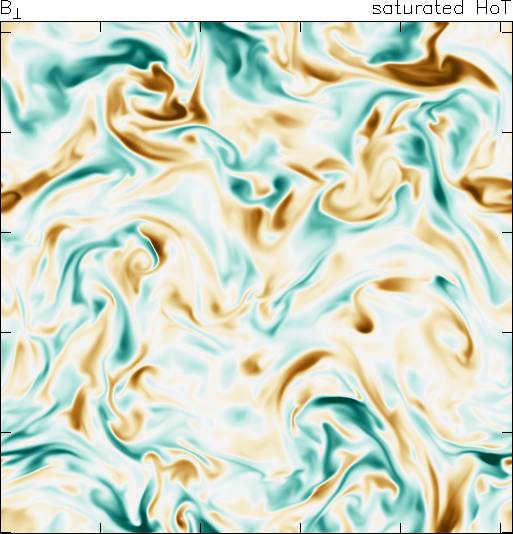}&
\includegraphics[height=.32\linewidth]{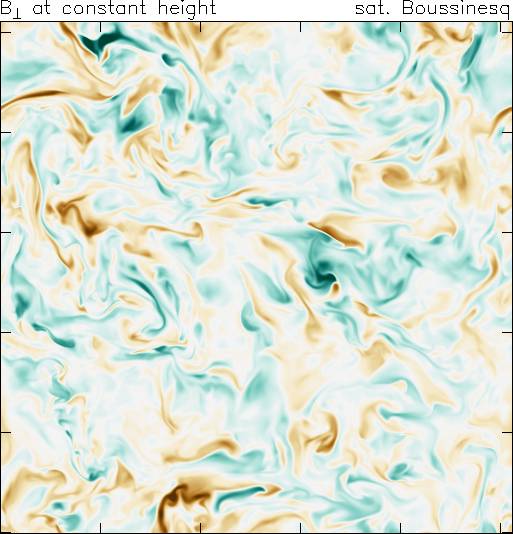}&
\includegraphics[height=.32\linewidth]{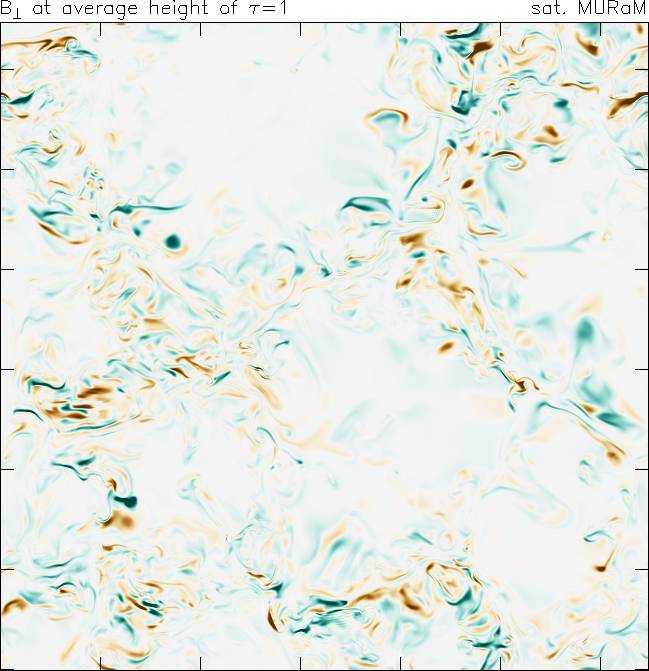}\\
\includegraphics[height=.32\linewidth]{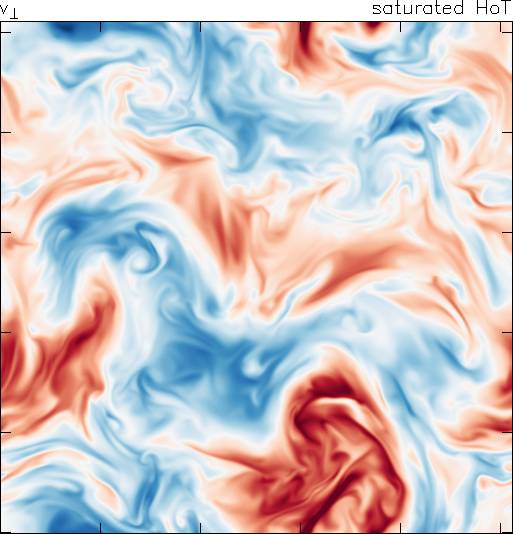}&
\includegraphics[height=.32\linewidth]{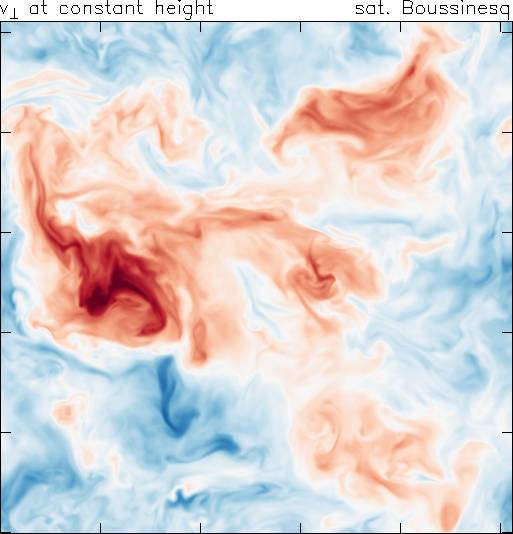}&
\includegraphics[height=.32\linewidth]{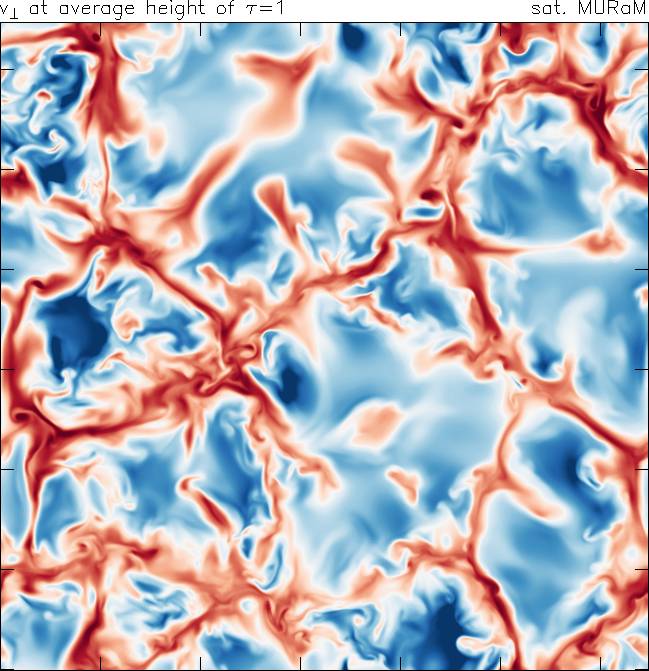}
\end{tabular}
\caption{Maps of the vertical components of magnetic field (upper row)
and fluid velocity (lower row) on horizontal cuts through each
simulation box, taken from snapshots in the saturated dynamo state.
The two magnetic polarities are shaded in turquoise and brown; white
corresponds to zero vertical field. In the velocity maps, downflows
are shaded red and upflows blue. Note that the arbitrary ratio
of driving scale to box size is a factor of 3 smaller for \MURaM.}
\label{fig:slices}
\end{figure*}

In the saturated state, the ratio of total magnetic to total kinetic
energy is 0.41 for HoT, 0.36 for BC, and 0.026 for \MURaM. The
small-scale dynamo is more efficient in the homogeneous cases HoT and
BC. In the stratified solar case simulated with \MURaM, the
restriction of dynamo action to the downflow lanes and the losses due
to downflows leaving the computational box limit the level of magnetic
energy to a few percent of the kinetic energy.
Figure~\ref{fig:slices} illustrates the structure of the velocity
field and the dynamo-generated magnetic field in the saturated states
of the three simulations. Shown are maps of the vertical magnetic
field and velocity components on horizontal cuts.  In all three cases,
the magnetic field exhibits the typical folded structures arising from
small-scale dynamo action, with elongated unipolar features and rapid
polarity reversals in the transversal direction, often on a resistive
spatial scale. While the field structures are fairly homogeneously
distributed in the HoT and BC cases, the \MURaM simulation shows an
intermittent structure with extended patches almost devoid of field.
This structure results from the up-down asymmetry of convection in a
stratified medium: the convective upflows expand heavily in the
horizontal directions and thus expel the magnetic flux. Thus in the
\MURaM simulation, magnetic field generation takes place mainly in the
vicinity of the narrow turbulent downflows.  The narrowness of these
downflows and the size of the convection cells relative to the
arbitrarily chosen size of the simulation box contribute to a
smaller appearance of the magnetic field structures for \MURaM
presented in Figure~\ref{fig:slices}.

\begin{figure*}[t]
\begin{tabular}{c@{\hskip1pt}c@{\hskip1pt}c}
\includegraphics[height=.51\linewidth]{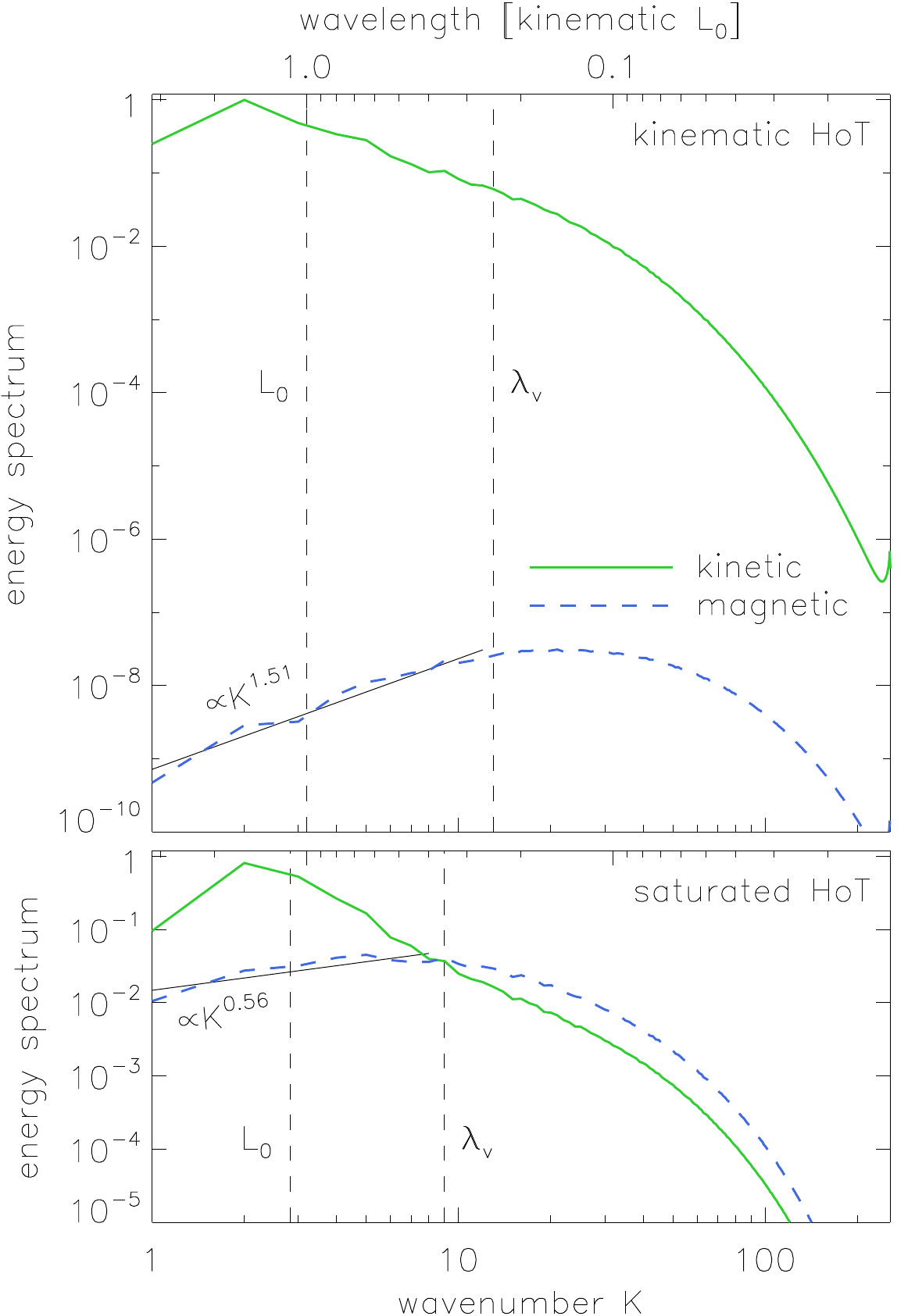}&
\includegraphics[height=.51\linewidth]{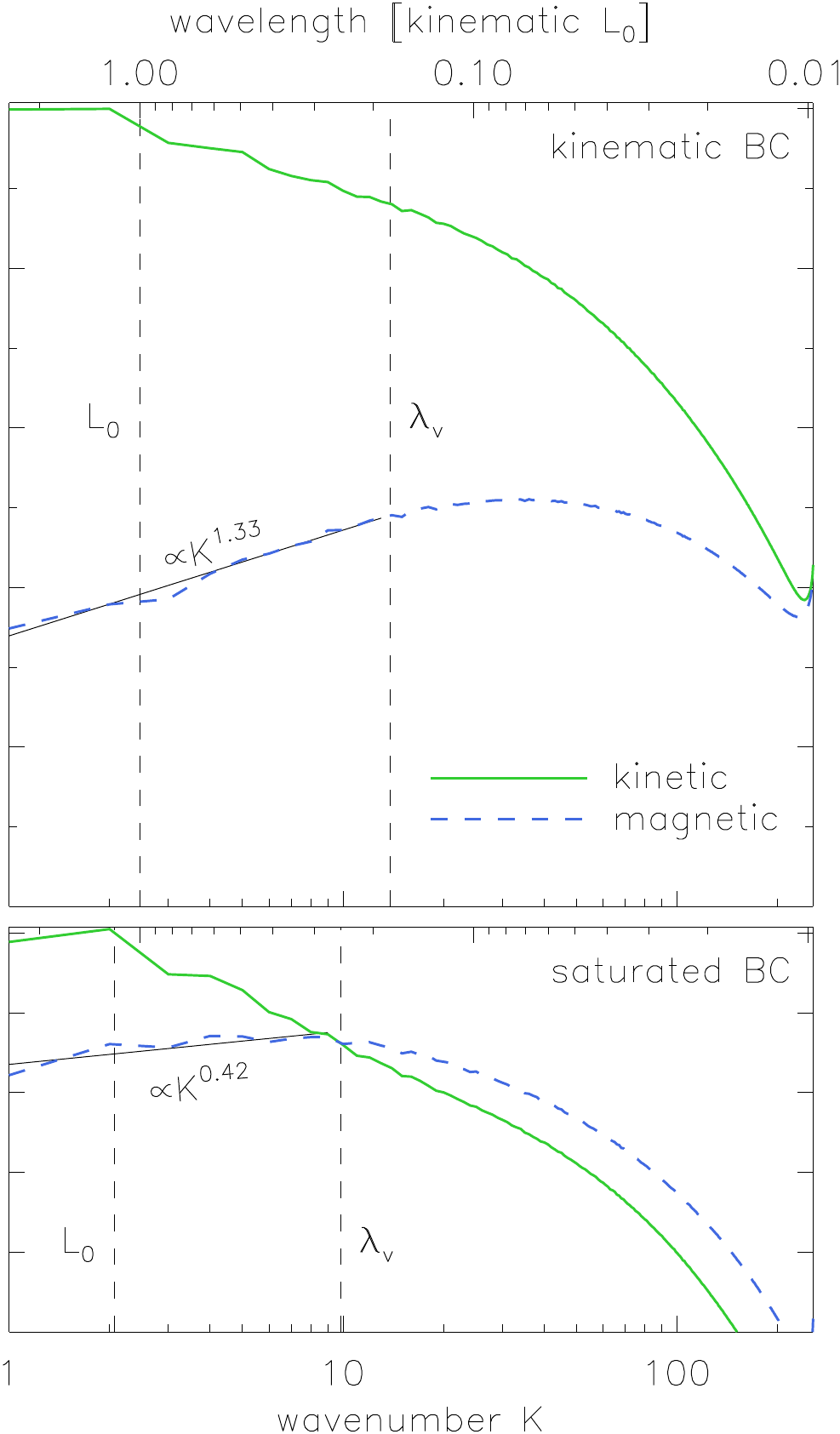}&
\includegraphics[height=.51\linewidth]{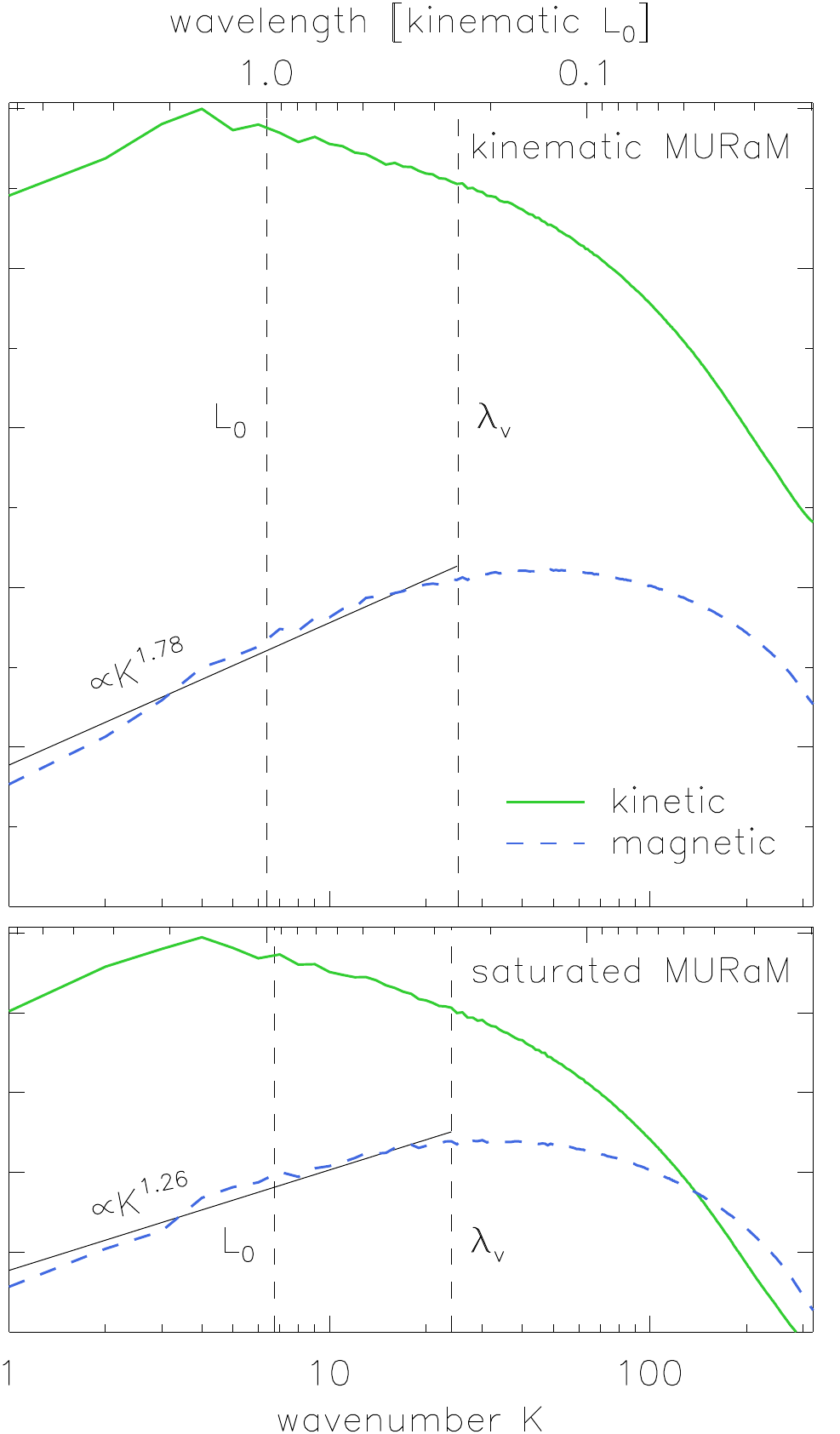}\\
\end{tabular}
\caption{Kinetic and magnetic energy spectra in the kinematic (upper)
  and saturated (lower) states of the homogeneous turbulence (HoT)
  simulation (left), the Boussinesq convection (BC) simulation
  (center), and the solar convection (\MURaM) simulation (right).}
\label{fig:enspecs}
\end{figure*}

The greater separation between the driving scale and the box
size can be quantified in the kinetic and magnetic energy spectra for
the simulations shown in Fig.~\ref{fig:enspecs}. Vertical lines
indicate the integral scale for the turbulent motions,
\begin{equation}
    L_0 = \left. \int_0^\infty k^{-1} \EV(k) \,\dek
    \middle/ \int_0^\infty \EV(k) \,\dek, \right. 
\end{equation}
and the Taylor microscale $\lambv$,
\begin{equation}
    \lambv^2 = \frac{\mean{v^2}}{\mean{\omega^2}} 
       = \left. \int_0^\infty \EV(k) \,\dek
    \middle/ \int_0^\infty k^2 \EV(k) \,\dek. \right.
\end{equation}
These scales signify the approximate beginning and end of the inertial
range for a hydrodynamic or kinematic state.  Both scales
occur at higher wavenumbers for \MURaM.

\((L_0/\lambv)^2 \), which is proportional to the effective Reynolds
number, is $31.5$ in the kinematic and $22.5$ in the saturated state
of the Boussinesq simulation and $16.5$ ($10.1$) in the HoT
simulation. The corresponding values for \MURaM are $15.7$ and $12.7$,
respectively.  The spectra of all three dynamos are similar.  In
Fig.~\ref{fig:enspecs}, the magnetic energy spectra exhibit power laws
with positive exponents at small wavenumbers, peaking beyond $\lambv$
in the kinematic state. The omni-directional spectra for \MURaM are
calculated employing a Tukey window \citep{1978Harris} in the vertical
direction.  The window corresponds to a high-resolution/low-dynamic
range apodization. This apodization increases spectral leakage from
much stronger disparate frequencies such as the peak of kinetic energy
at spatial frequency $L_0^{-1}$. Using a
high-dynamic-range/low-resolution window such as the Nuttall window
leads to similar spectral indices. In the stratified \MURaM
simulation, the anisotropic flow complicates any interpretation of the
power law exponents. In the saturated state, magnetic energy dominates
over kinetic energy for the largest wavenumbers and the peak of the
spectrum moves to lower wavenumbers.

\subsection{Shell-to-Shell Transfer}
\label{sec:s2s}

\begin{figure}[tp]
    \includegraphics[width=\linewidth]{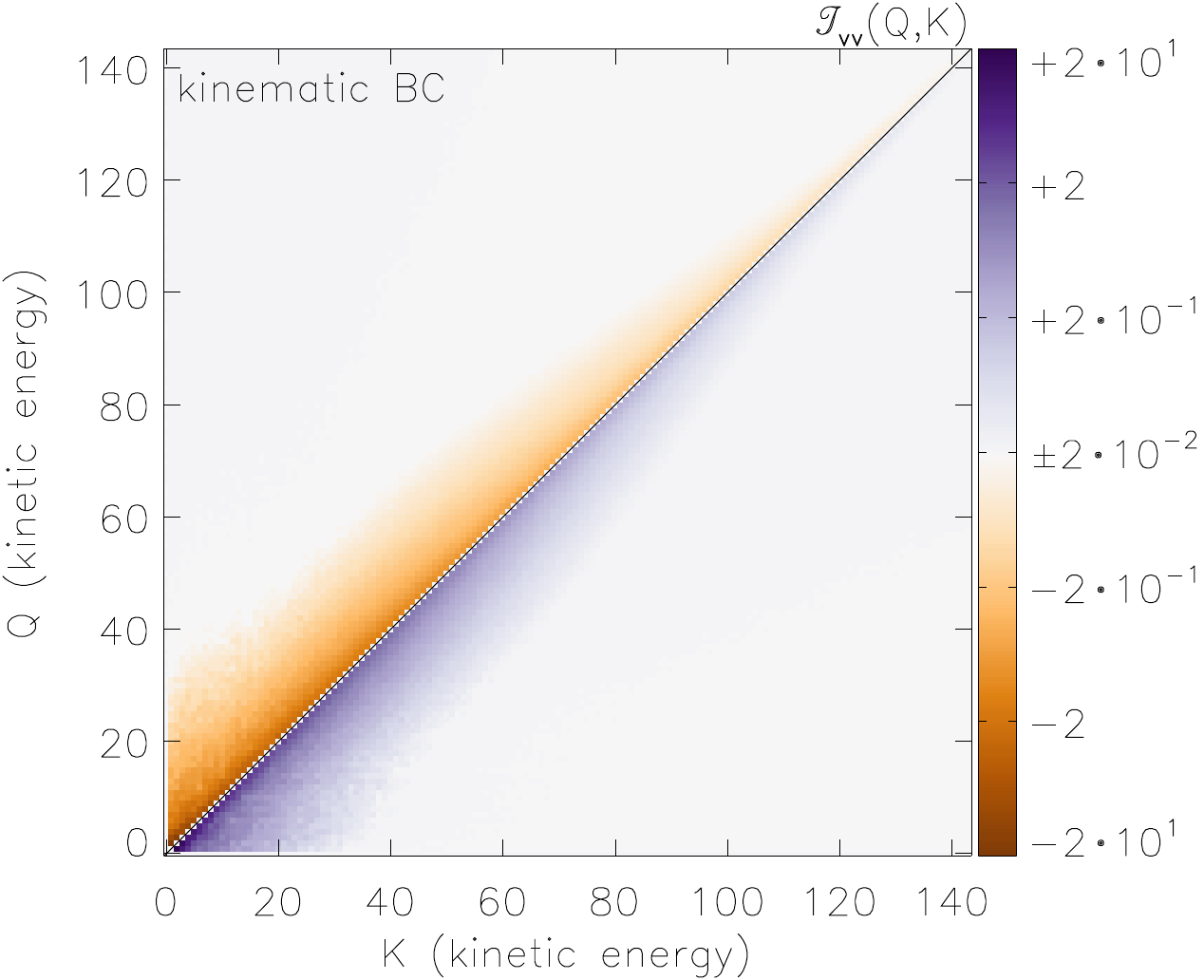} \\
    \includegraphics[width=\linewidth]{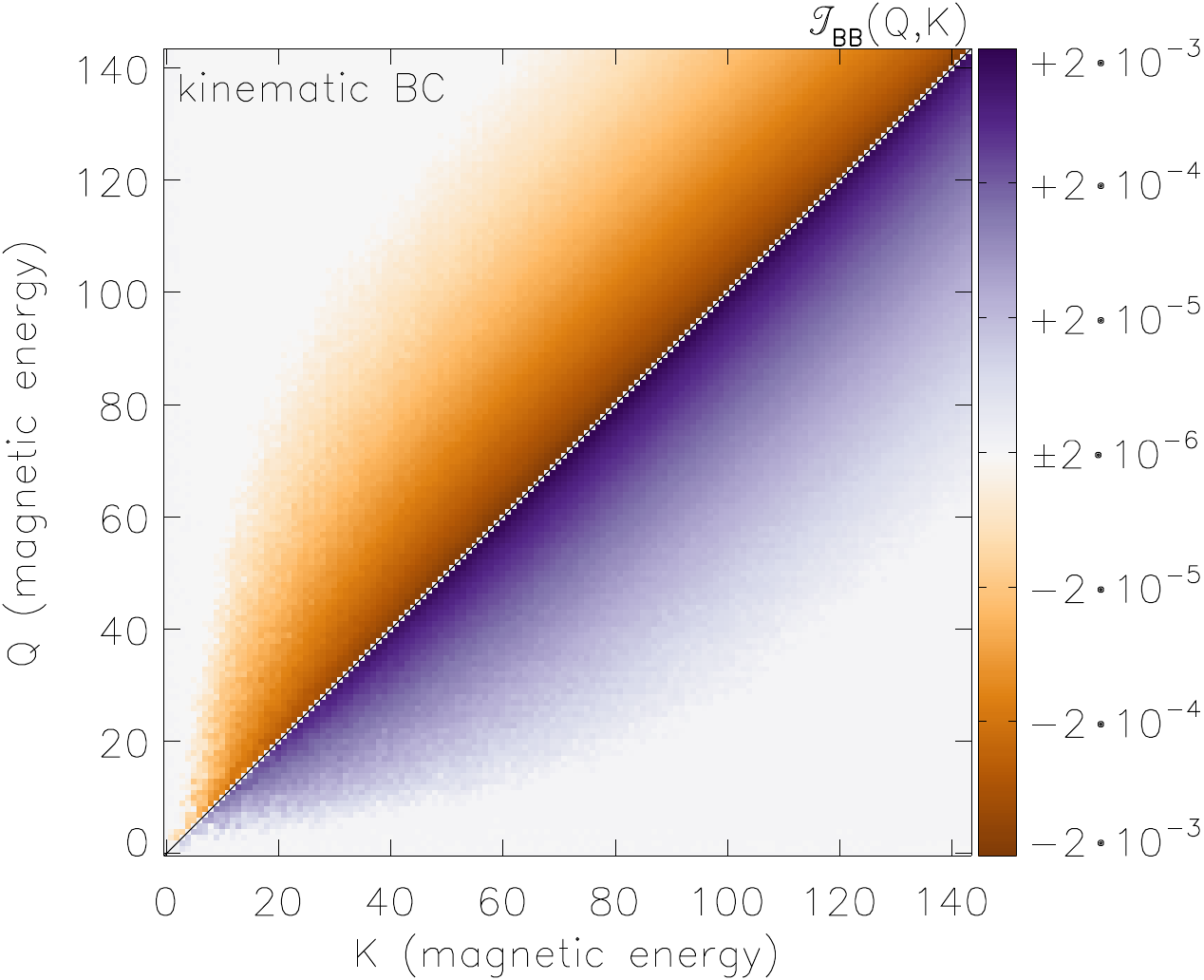} 
\caption{Shell-to-shell transfer from wavenumber $Q$ to wavenumber $K$ in the
kinematic state of the  BC simulation.  Top: kinetic to kinetic energy.
Bottom: magnetic to magnetic energy.}
\label{fig:uubbkin}
\end{figure}

In order to compare the dynamo mechanism in our simulations, we need to
look beyond visual and spectral similarities and measure in detail the
generation of magnetic energy.  To do this, we employ the shell-to-shell
analysis described in Section~\ref{sec:transan}.

As an example of how to interpret the transfer functions \(\TT(Q,K)\),
Fig.~\ref{fig:uubbkin} shows the cascade of kinetic and magnetic energy
towards larger wavenumbers in the kinematic phase of the BC dynamo. In
all of the plots, the transfer functions have been normalized by the
maximum value of $\TVBT$ in the saturated state for each type of
simulation. The normalization allows straightforward comparison of the
relative changes in $\TVBT$, which is important for understanding the
saturation mechanism of the dynamo. The normalization of all transfer
functions for each type of simulation is consistent; however the
absolute values of $\TVBT$ depend on the total amount of energy. Since
the total energies vary between the simulations, the difference in the
absolute magnitudes between simulations should not be compared.  Instead
we focus on qualitative comparisons and relative changes, both of which
are not affected by the normalization.  As the values of the different
transfer functions vary strongly, the 2D plots have been constructed to
extend over three orders of magnitude of positive and negative
values. The absolute upper limit of the color map corresponds to the
absolute maximum value of the represented $\TT$. 

In the top panel of Fig.~\ref{fig:uubbkin}, the transfer of kinetic
energy from one shell to another, $\TVV(Q,K)$, is positive (indicated
by purple color) for $Q\la K$, i.e., shell $K$ receives energy from
slightly smaller wavenumbers $Q$. $\TVV(Q,K)$ is negative (orange) for
$Q\ga K$, i.e., shell $K$ loses energy to slightly larger wavenumbers
$Q$. Kinetic energy is transferred to larger wavenumbers as part of
the local direct cascade of energy.  This is seen also in the transfer
of magnetic energy from one shell to another $\TBB(Q,K)$ (bottom
panel).  This picture is the same for both homogeneous (not shown) and
Boussinesq convection and for both the kinematic and saturated (not
shown) states \citep[also compare with Fig.~3 of][]{2005Mininni}.

\begin{figure*}[t]
\begin{tabular}{r@{\hskip7pt}l}
    \includegraphics[height=.39\linewidth]{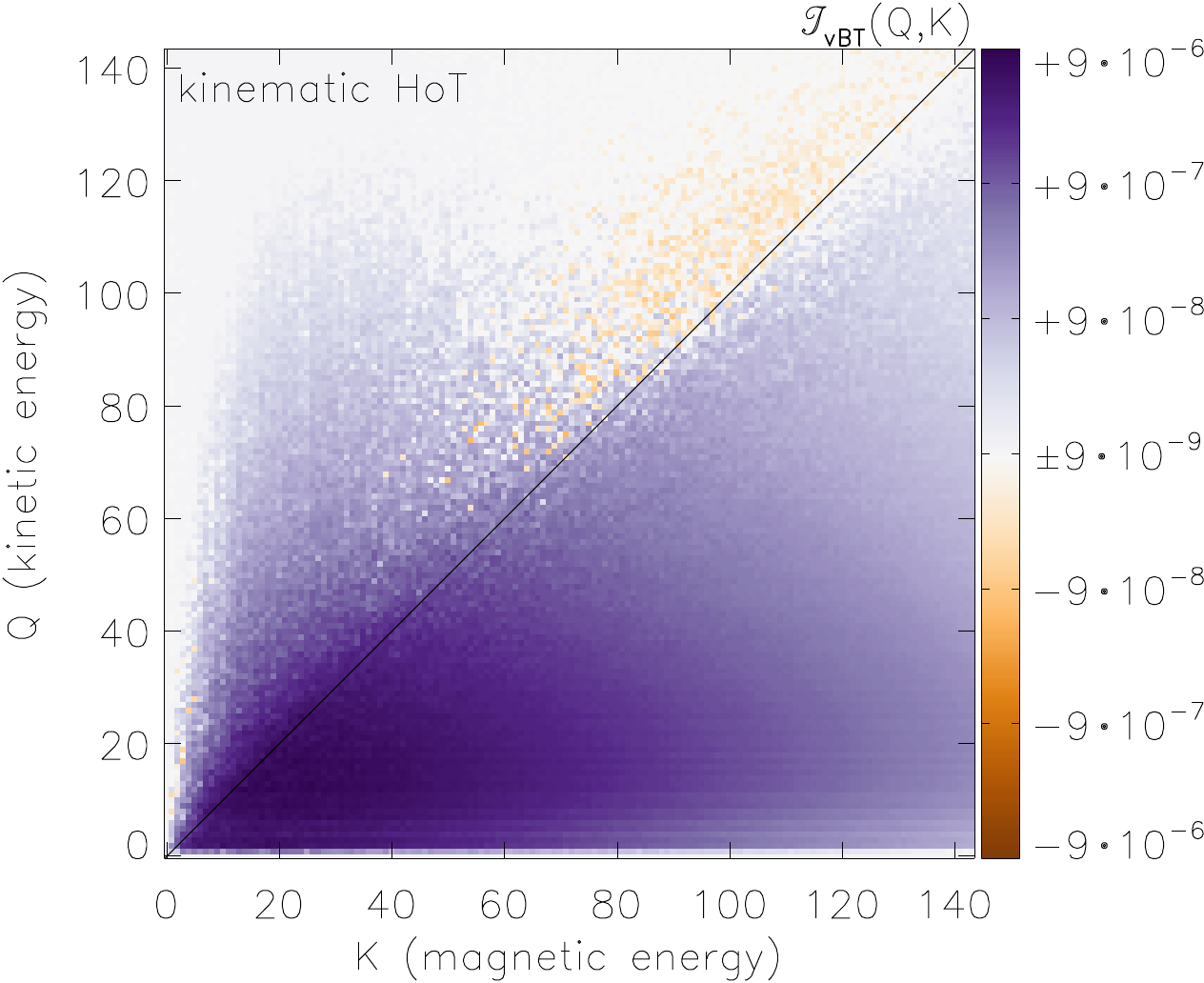} &
    \includegraphics[height=.39\linewidth]{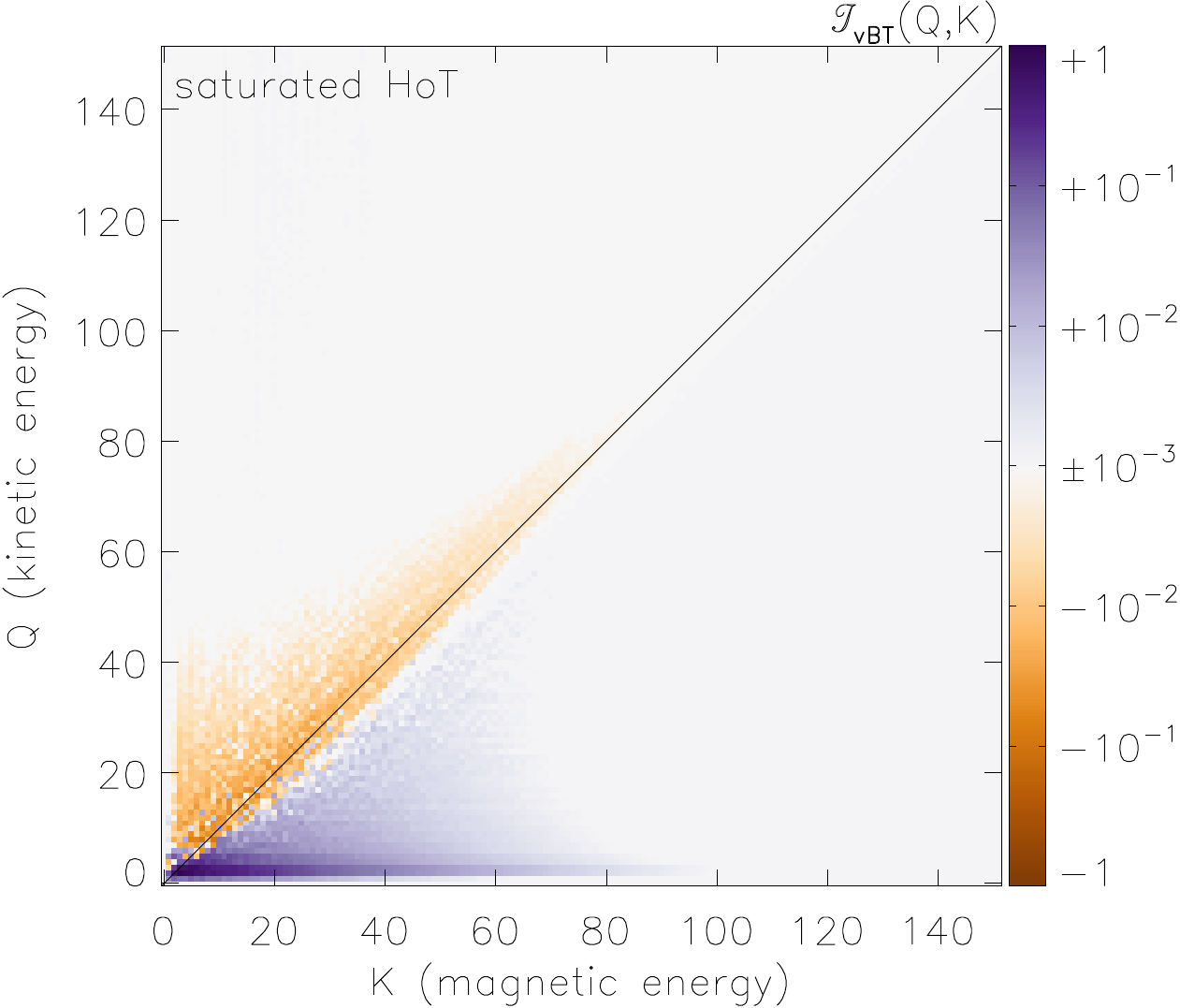} \\
    \includegraphics[height=.39\linewidth]{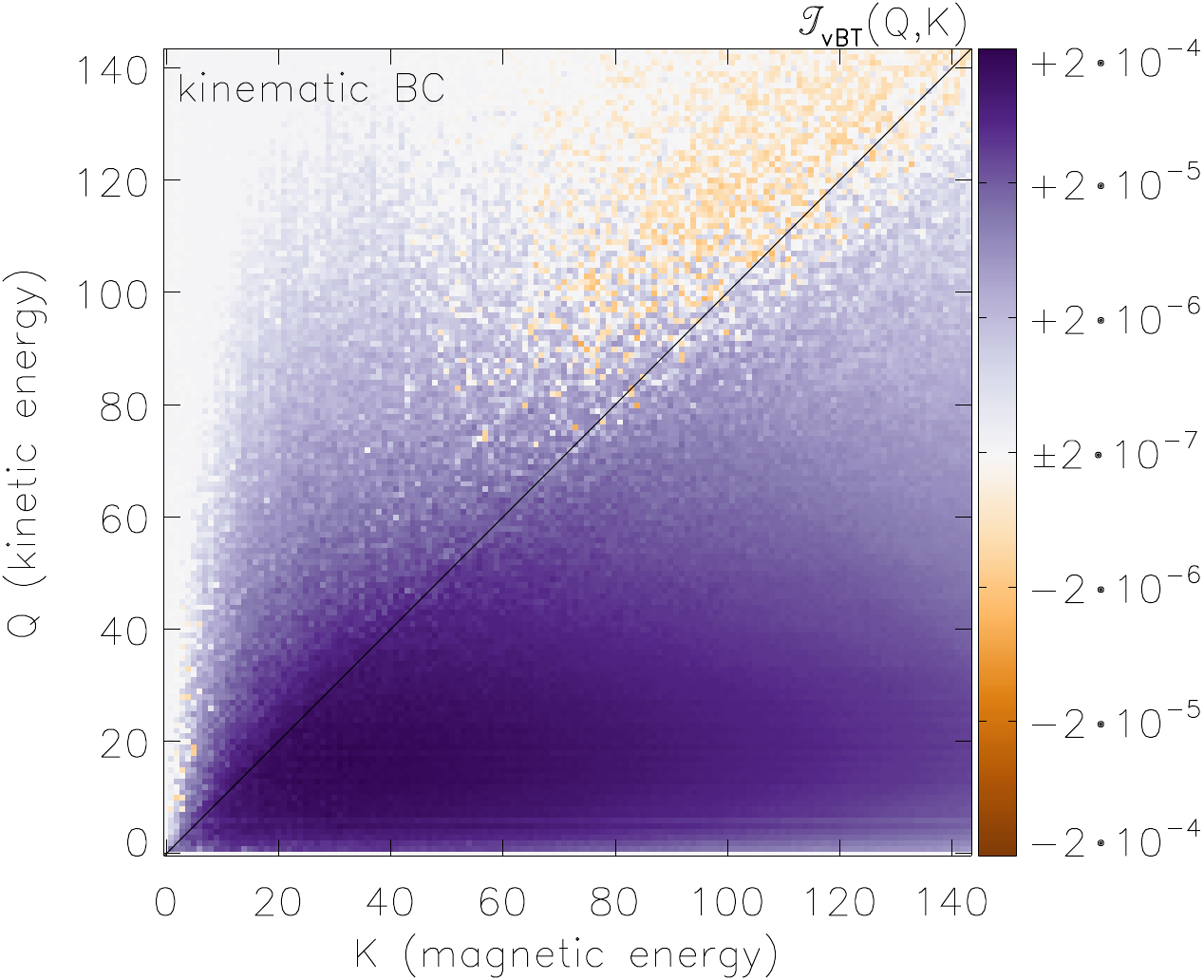}&
    \includegraphics[height=.39\linewidth]{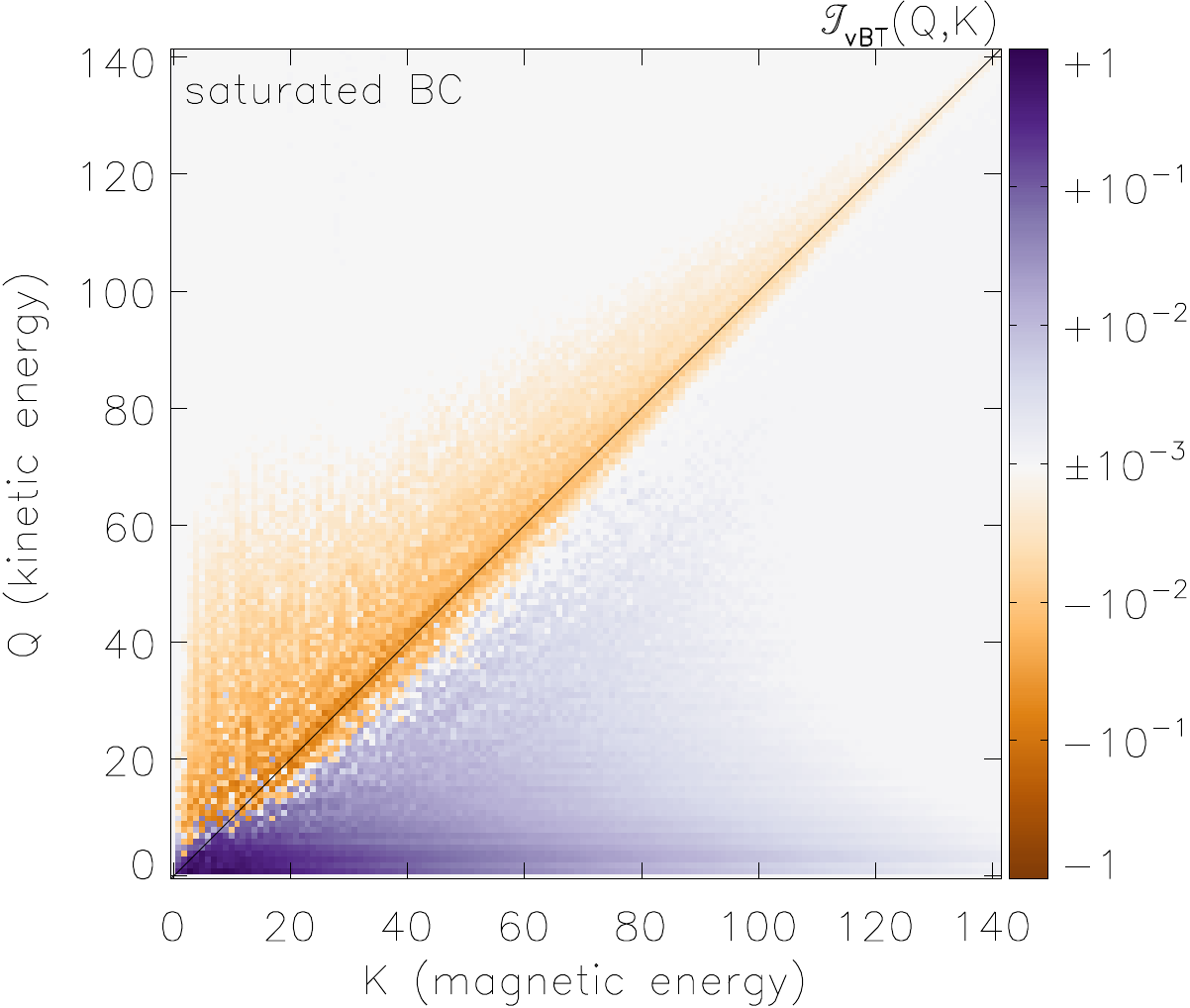}\\
    \includegraphics[height=.39\linewidth]{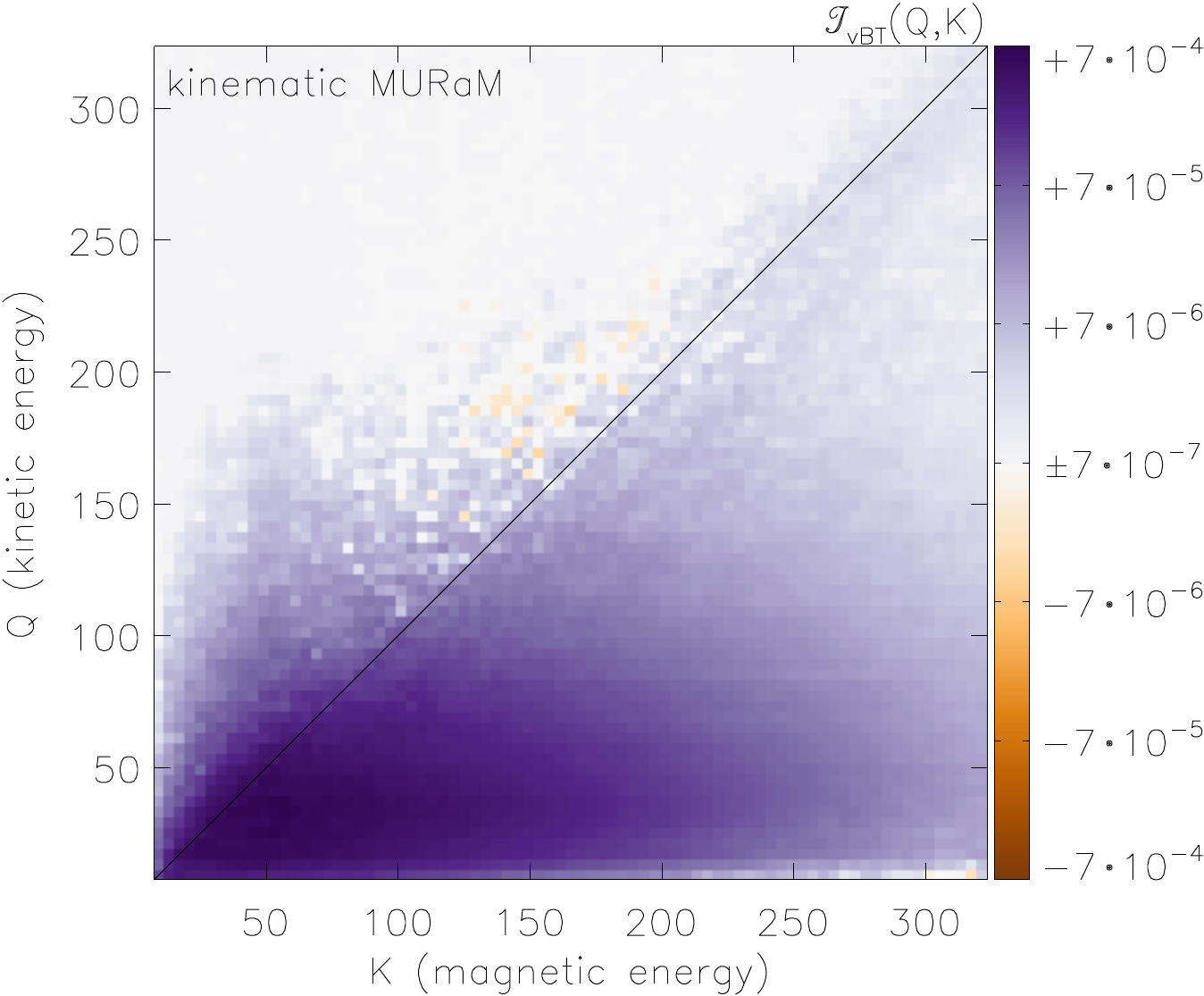} &
    \includegraphics[height=.39\linewidth]{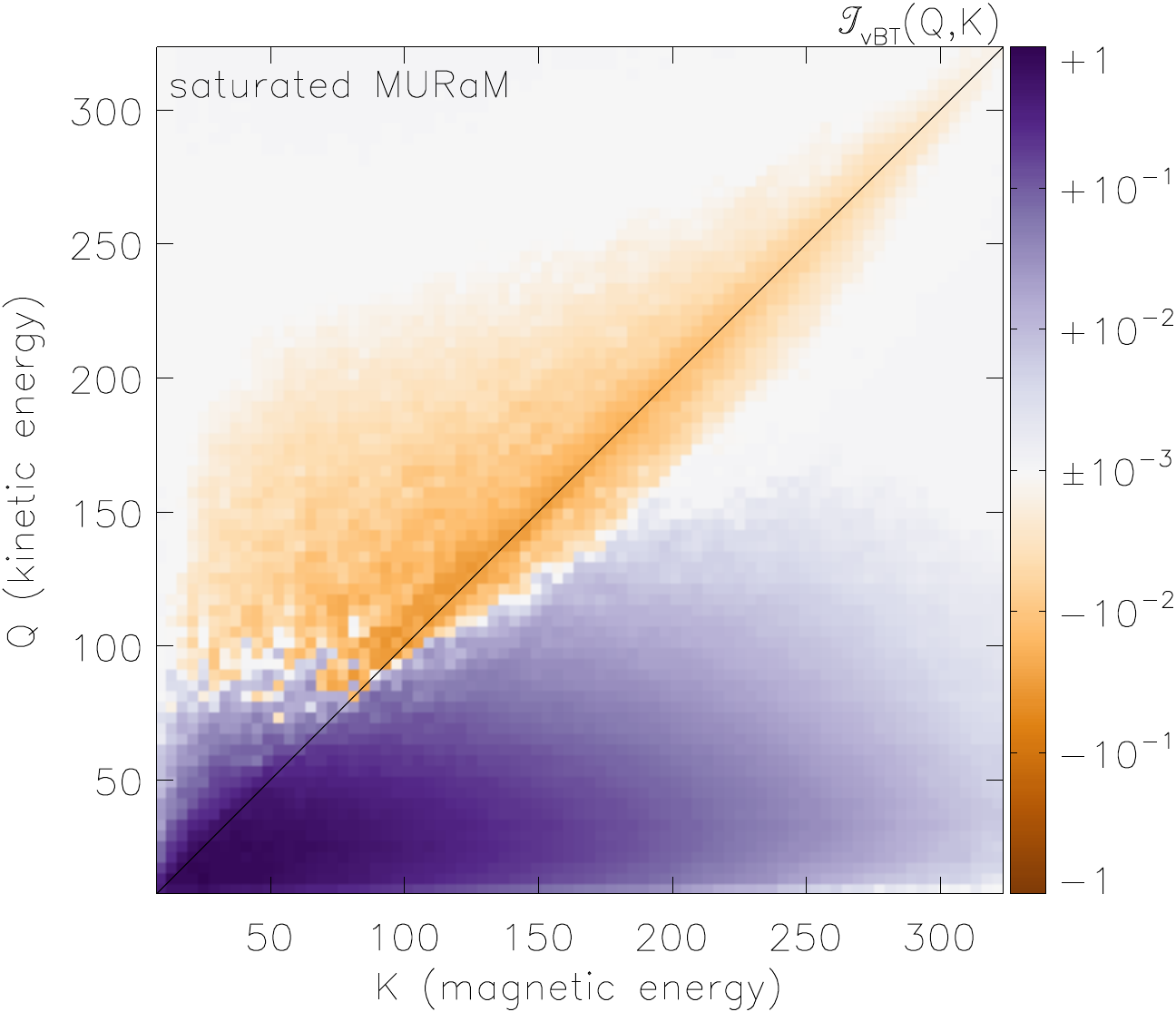} \\
\end{tabular}
\caption{ Shell-to-shell transfers for homogeneous turbulence (HoT, top
panels), Boussinesq convection (BC, middle panels) and solar convection
(\MURaM, bottom panels).  Shown are the transfers of kinetic energy at
wavenumber shell $Q$ to magnetic energy at shell $K$, $\TVBT(Q,K)$, in
the kinematic (left panels) and saturated states (right panels).  Within
each of the three types of simulations, a given set of arbitrary units
is employed.}
\label{fig:ubbou}
\end{figure*}

\begin{figure*}[t]
\begin{tabular}{r@{\hskip7pt}l}
    \includegraphics[height=.34\linewidth]{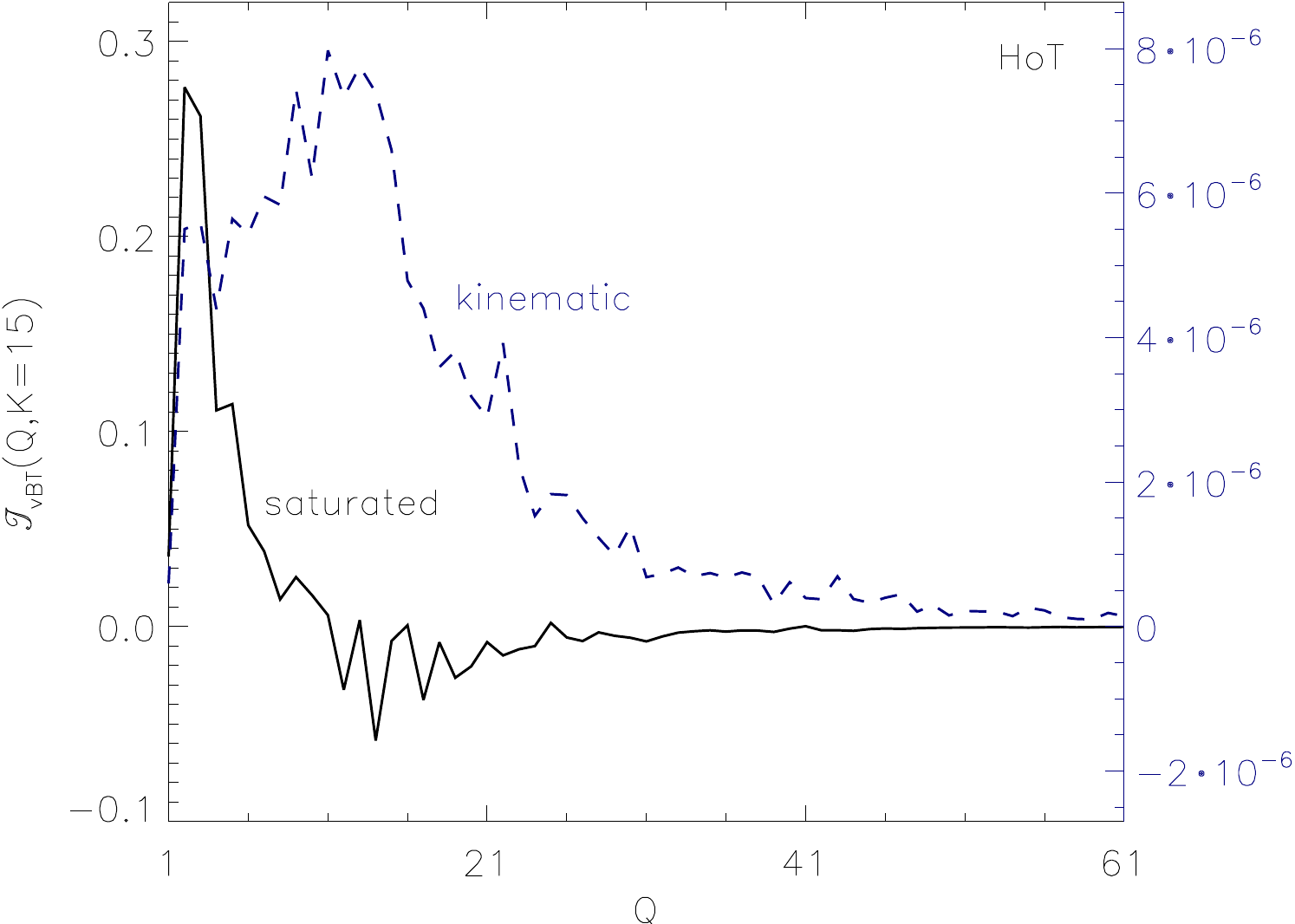} &
    \includegraphics[height=.34\linewidth]{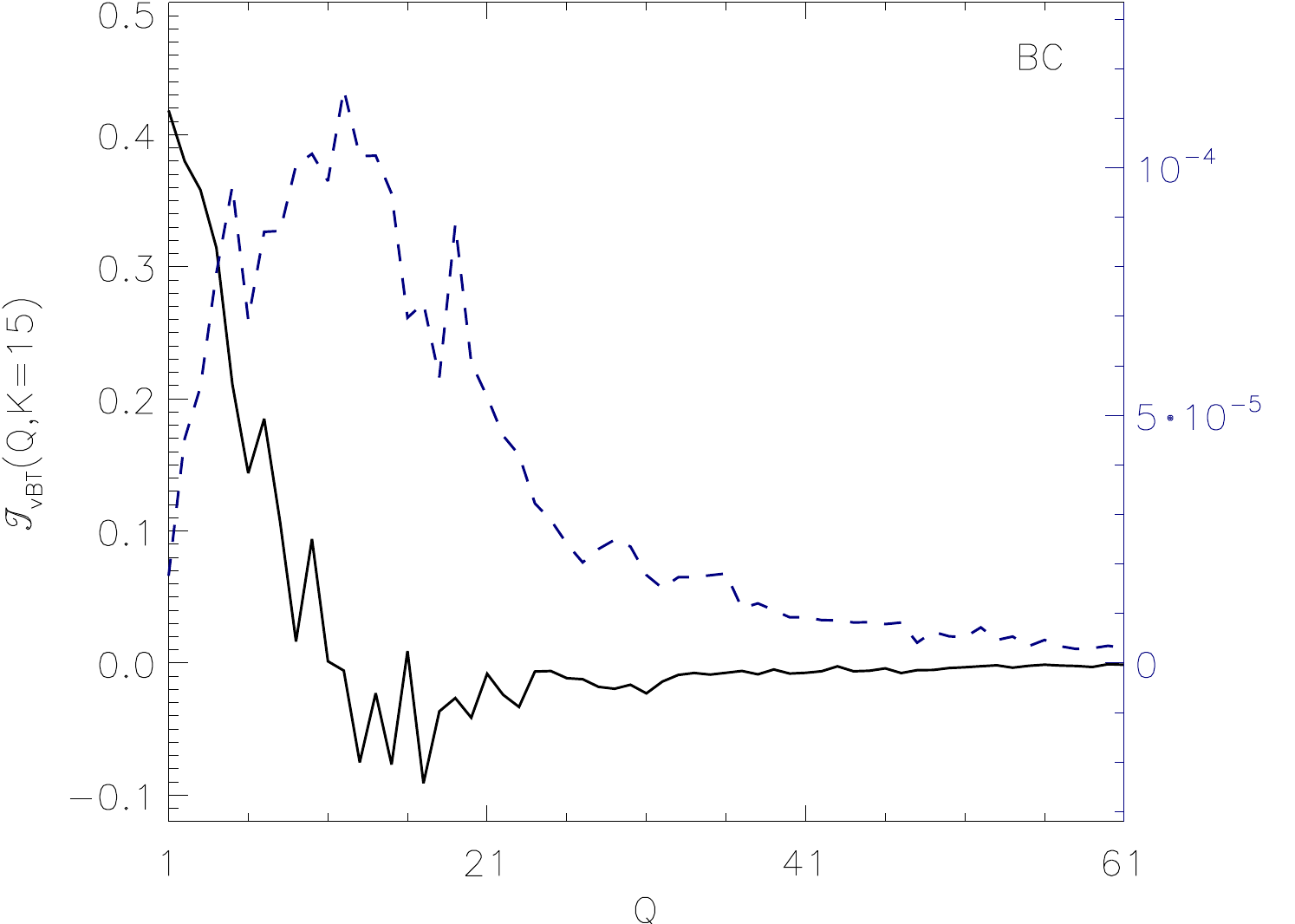} \\
    \includegraphics[height=.34\linewidth]{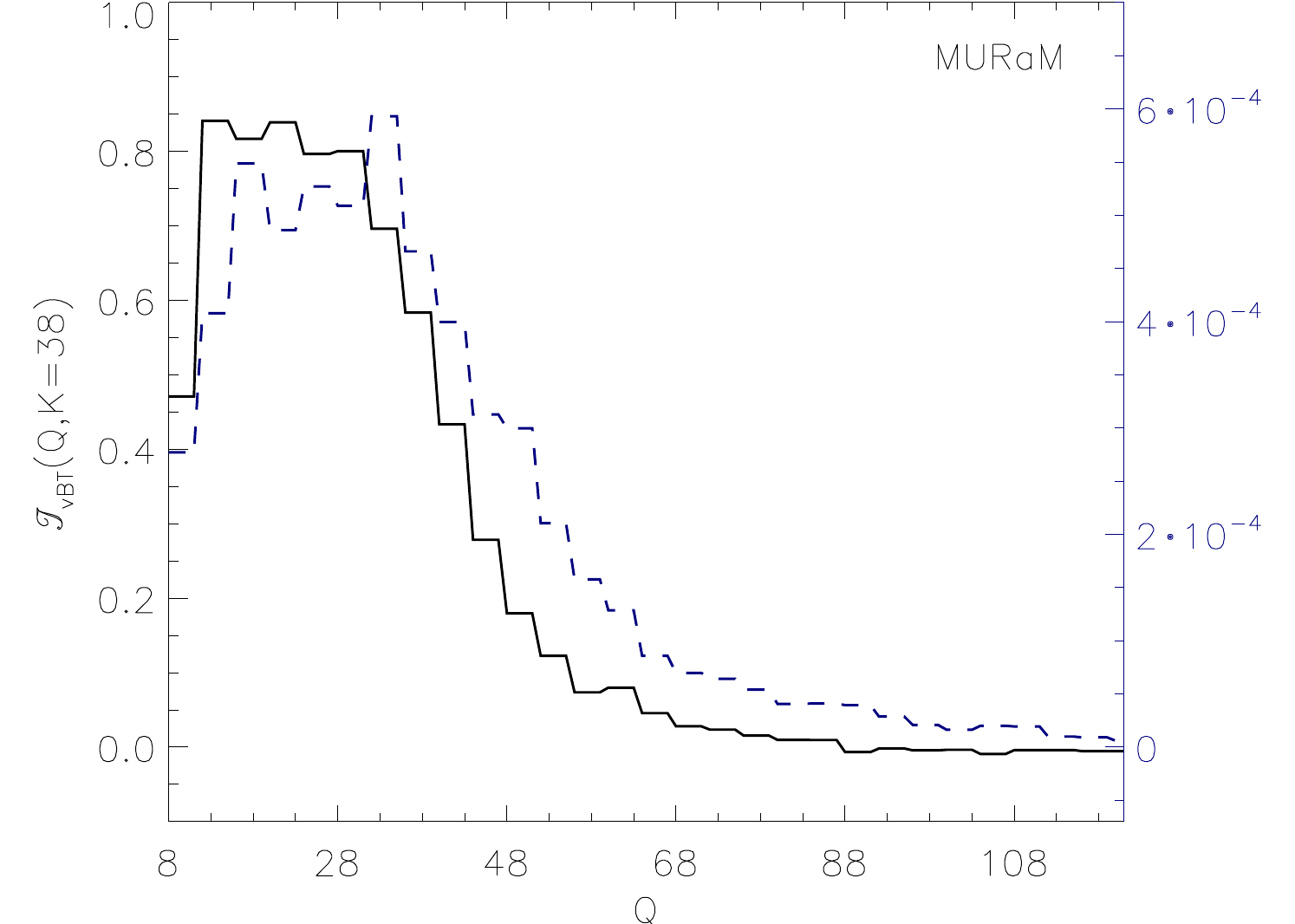}&
    \includegraphics[height=.34\linewidth]{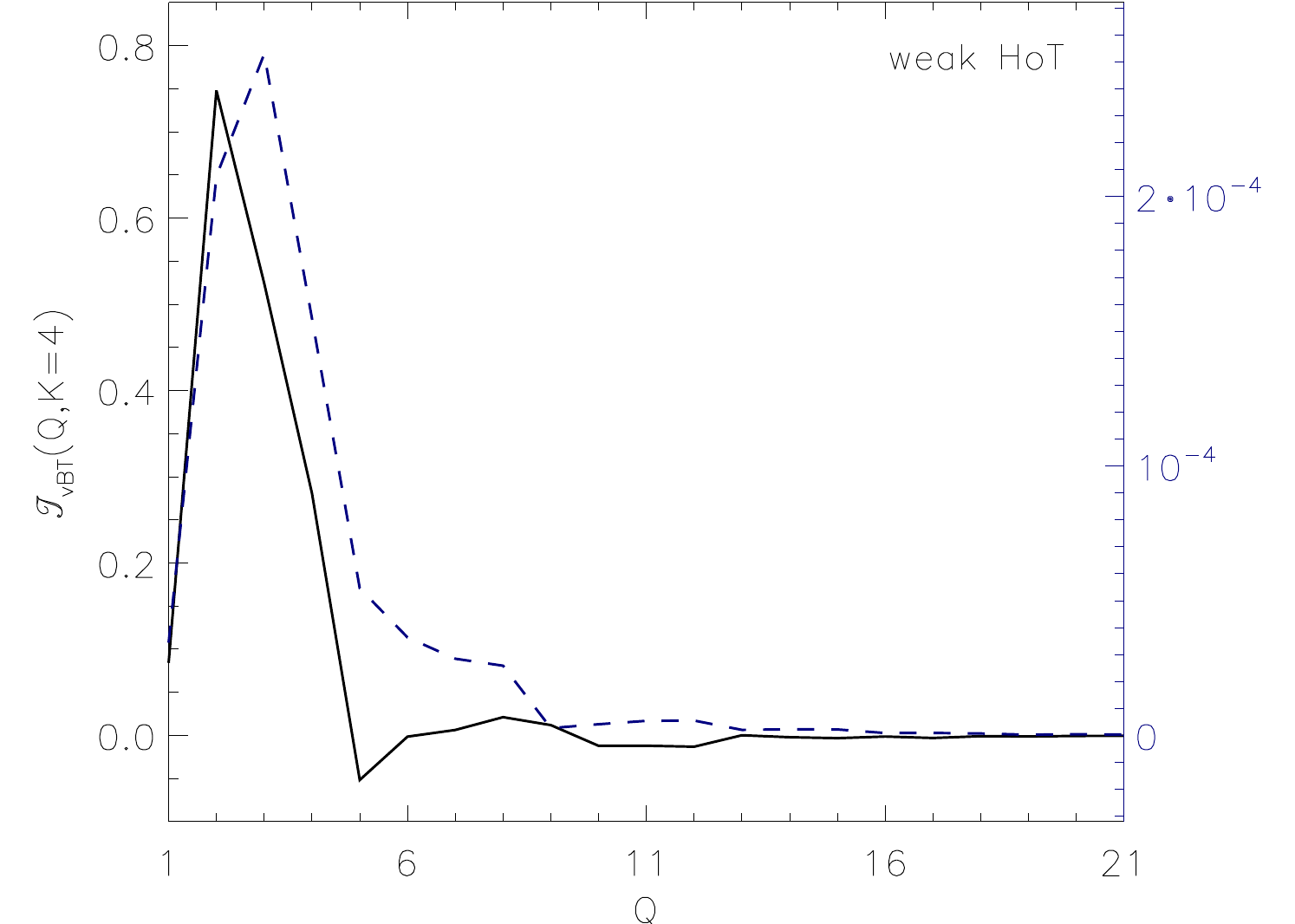}\\
\end{tabular}
\caption{Cuts at constant $K\approx1.3/\lambv$ through the 2D maps in
Figure~\ref{fig:ubbou} for the homogeneous turbulence (HoT) simulation (top
left, $K=16$), the Boussinesq convection (BC) simulation (top right, $K=16$),
the solar convection (\MURaM) simulation (bottom left, $K=38$) and the weak
homogeneous turbulence simulation ($\Emag=10\%\Ekin$, $\Rey\approx270$, bottom
right, $K=5$).  Solid lines correspond to the saturated state (left vertical
axis) and dashed lines to the kinematic state (right vertical axis).}
\label{fig:ubcut}
\end{figure*}

The kinematic dynamo mechanism is analyzed by measuring $\TVBT$, shown
in panels on the left-hand side of Fig.~\ref{fig:ubbou}.  This transfer is
positive for all $Q<K$, not just for nearby shells as in the cascade
process in Fig.~\ref{fig:uubbkin}. All wavenumbers $Q<K$ of the fluid
motion drive magnetic energy in a given shell $K$.  This is the same
as seen for dynamos driven by a mean flow \citep{2005Mininni}. Here,
we see the same kinematic dynamo mechanism in all 3 simulations.
At very large $Q>K$ corresponding to dissipative wavenumbers,
there is less negative transfer in the \MURaM case than in the other
two cases, where it is already very weak.  We conjecture that this
difference is due to the differing treatment of viscosity in the
codes.

The saturation of the dynamo results from the suppression of the
velocity fluctuations at larger wavenumber by the back-reaction of the
magnetic field on the flow through the Lorentz force. In the panels on
the right-hand side of Figure~\ref{fig:ubbou} the transfer $Q\ga K$ is
negative (indicated by an orange color). This generalizes the result
of \citet{2005Mininni} not only to random forcing in homogeneous
turbulence, but also to incompressible convection and to realistic
solar-like stratified convection.  The wavenumber range where
$\TVBT<0$ coincides with the range at which the magnetic energy
spectrum dominates the kinetic energy spectrum (cf.
Fig.~\ref{fig:enspecs}).  In \MURaM, which has a lower saturated ratio
$\Emag/\Ekin\approx 2.6\%$ and a larger scale separation from
the box size, this point occurs at significantly larger wavenumbers
than for the other two simulations.

In the saturated state, there is a transfer of kinetic energy to
higher wavenumber ($K>Q$) magnetic energy similar to that seen in the
kinematic state. However, the transfer is more localized towards the
small wavenumber $Q$ end of the kinetic energy spectrum
\citep{Alexakis:etal:2005,2005Mininni,2006Carati}. For example, in the
HoT simulation, the driving wavenumber for the fluid motions,
$Q\approx3$, is the dominant source for transfers to all $K$ in the
saturated state.

Figure~\ref{fig:ubcut} presents vertical cuts at constant $K$ through
the 2D maps of $\TVBT(Q,K)$ shown in Figure~\ref{fig:ubbou}. These
slices allow us to quantify the changes between the kinematic and the
saturation regimes.  For the \MURaM dynamo, both the shift towards
smaller $Q$ and the amount of negative transfer for $Q\ga K$ is less
dramatic than for the homogeneous and Boussinesq dynamos.  This is
reasonable because the \MURaM dynamo is only slightly super-critical
and thus the dynamo action is an order of magnitude weaker than in the
other two simulations.  If we were able to conduct the analysis for
much larger $\Reym$, the effect would become more pronounced.
An additional HoT simulation performed in a regime similar to the
\MURaM dynamo confirms this interpretation.  This additional weak HoT
simulation was conducted at $\Rey=\Reym\approx270$ and
$\Emag=10\%~\Ekin$ in the saturated state. The \MURaM dynamo transfers
are shown next to similarly shaped transfers from the weak HoT in the
lower panels of Fig.~\ref{fig:ubcut}.

\begin{figure}[t]
    \includegraphics[width=\linewidth]{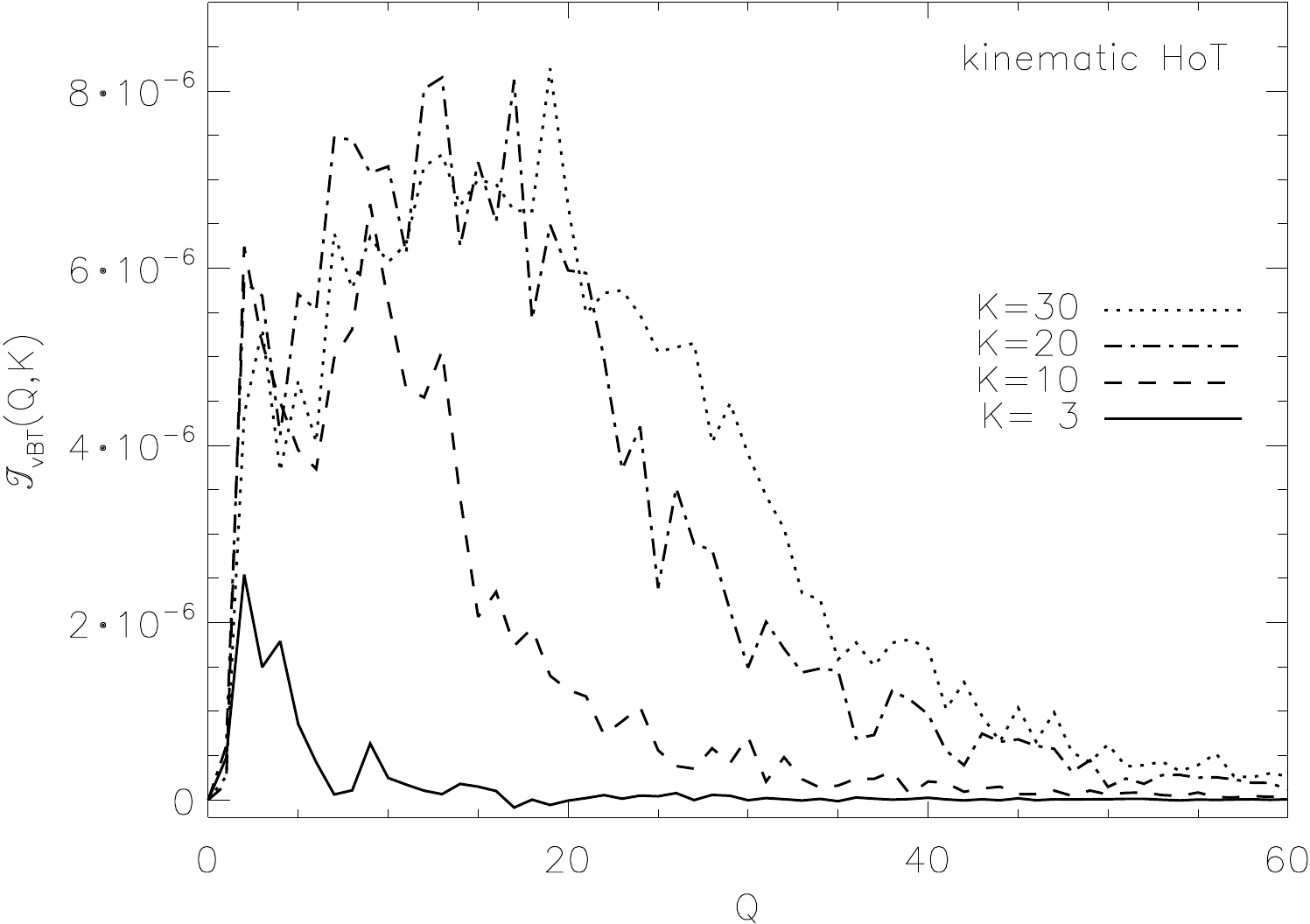}
\caption{Cuts at constant $K$ through the 2D map in
Figure~\ref{fig:ubbou} for the kinematic state of the homogeneous
turbulence (HoT) simulation for $K=3$ (solid), $K=10$ (dashed), $K=20$
(dot-dashed), and $K=30$ (dotted).}
\label{fig:local}
\end{figure}

In Fig.~\ref{fig:ubcut} a dominant peak at the driving
wavenumber is absent in the kinematic state. Such a peak
would be expected for a mean-flow driven incompressible dynamo
\citep[as seen in Fig.~6 of ][]{2005Mininni}.  For our randomly-forced
HoT simulation (Fig.~\ref{fig:local}), we find that the total transfer
to a given wavenumber $K$ increases with $K$, the transfer from
kinetic energy at $Q\approx3$ is no longer dominant for $K>3$, and the
total transfer to $K\approx3$ is dwarfed by the transfers to larger
$K$.  All three of these results present a view of randomly-forced
dynamos that differs strongly from mean-flow driven dynamos. In
particular, these results suggest that among small-scale turbulent
dynamos randomly-forced dynamos are more local than mean-flow driven
dynamos.  Turbulent fluctuations play a greater role, for the
kinematic phase, relative to integral-scale motions in the
randomly-forced case than in either mean-flow-driven case of
\citet{2005Mininni}.

\begin{figure}[t]
    \includegraphics[width=\linewidth]{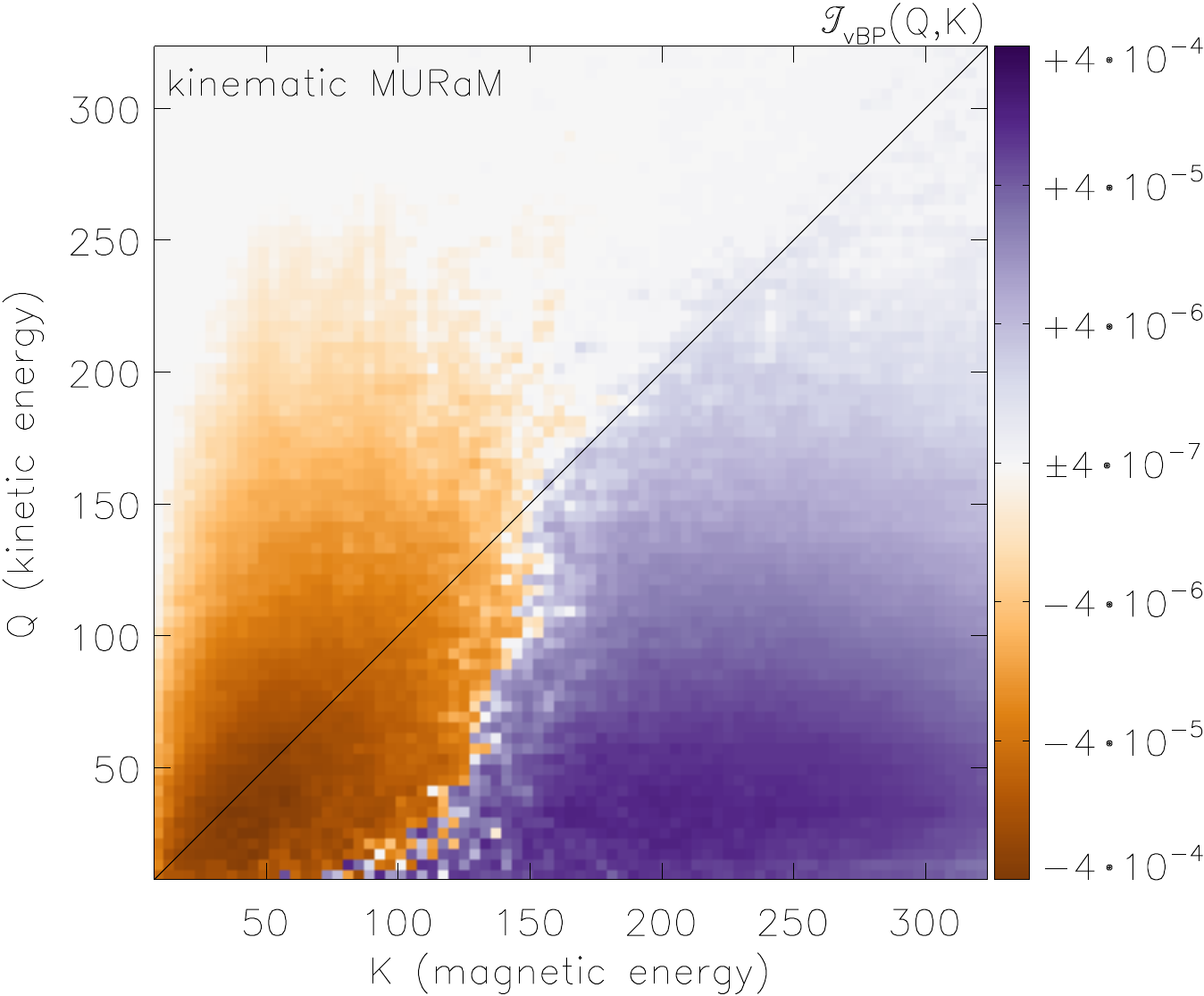} \\
    \includegraphics[width=\linewidth]{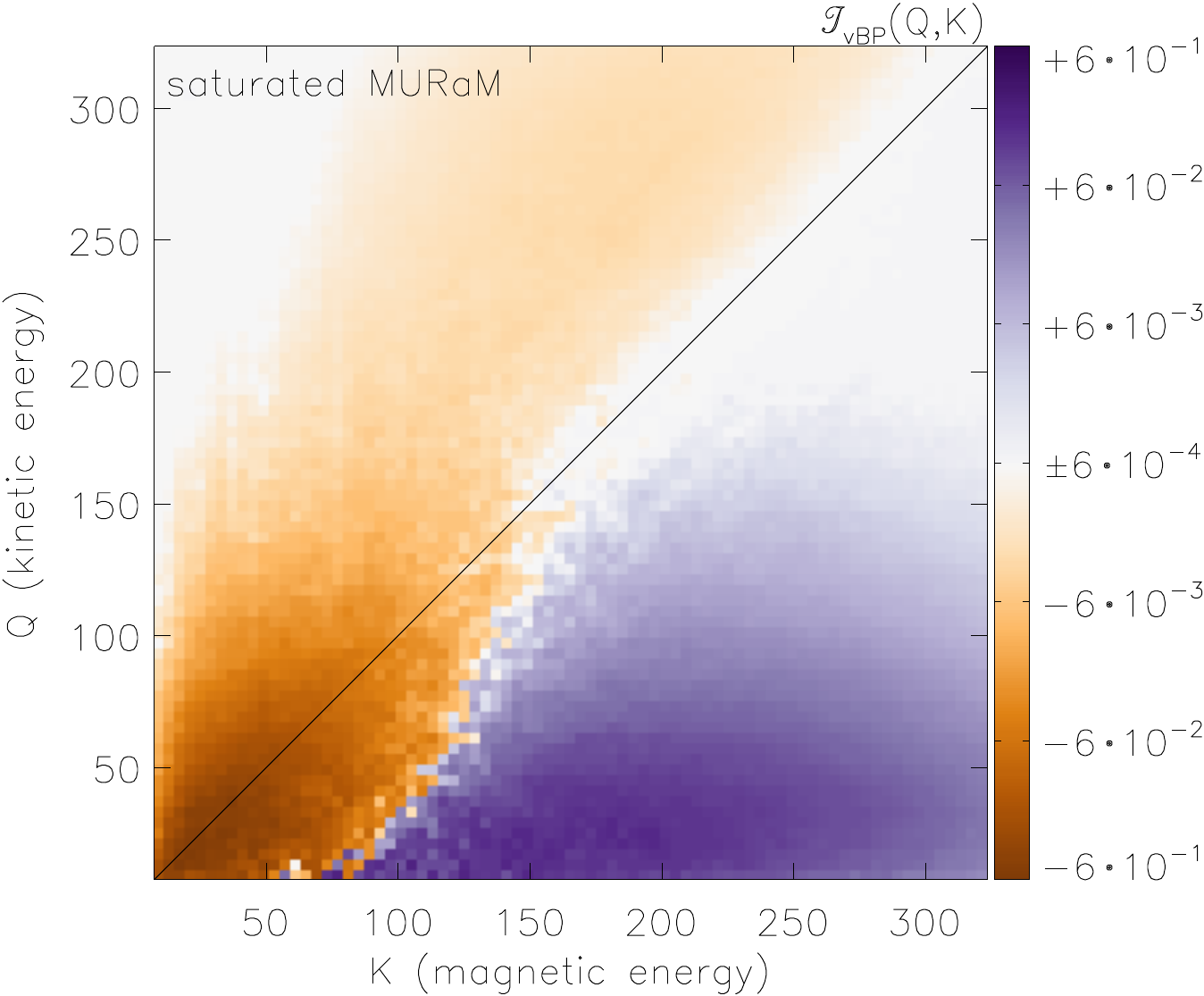} 
\caption{Shell-to-shell transfer of kinetic energy to magnetic energy in
    the solar (\MURaM) simulation through the magnetic pressure
    gradient, $\TVBP$, in the kinematic phase (upper panel) and in the
    saturated state (lower panel).}
\label{fig:s2s_mur}
\end{figure}

Work against magnetic pressure becomes involved in the transport of
magnetic energy from smaller to larger wavenumbers, a phenomenon often
referred to as the magnetic energy cascade.  For \MURaM this cascade
creates an imbalance resulting in an injection of 5\% of the magnetic
energy generated by the usual dynamo mechanism due to the magnetic
tension force \citep{Graham:etal:2010}. The shell-to-shell analysis of
the transfer $\TVBP$ is shown in Fig.~\ref{fig:s2s_mur}.  Net energy is
lost from small wavenumbers and deposited at larger wavenumbers.  The
transfer is not strongly dependent on the wavenumber $Q$ of the fluid
motions up to a break-over point, $K\approx100$.  Magnetic energy at
larger scales is expelled by the convection and compressed into the
downflows. Quantitatively, $\TVBP$ is stronger for smaller $Q$,
indicating the importance of motions at the convective scales for the
flux expulsion process.  At larger wavenumbers, magnetic energy is
generated by fluid motions at smaller wavenumber, e.g., through work
against the magnetic pressure, and lost to motions at larger wavenumbers,
such as viscously damped magnetosonic waves.

\section{DISCUSSION}
\label{sec:discussion}

In the three simulations presented here, turbulent flows are driven by
different mechanisms: random forcing (HoT), Boussinesq convection (BC) and
radiative cooling-driven convection (\MURaM).  The HoT and BC cases are very
similar in all aspects.  This includes the structure of the flow and magnetic
field (Fig.~\ref{fig:slices}), the energy spectra (Fig.~\ref{fig:enspecs}) and
the energy transfer spectra (Fig.~\ref{fig:ubbou}).  The similarity is not
surprising considering that the only essential difference between the
simulations is the phase information contained in the large-scale driving
function.

The outer-scale appearance of the \MURaM simulation is different from
the other two cases (Fig.~\ref{fig:slices}).  There is a strong
asymmetry between the upflowing and the downflowing plasma, with
relatively smooth upflows and narrow, highly turbulent downflows.
Also, the arbitrary ratio between the box size and the driving scale
is a factor 3 larger. In spectral space, this corresponds to a shift
towards higher wavenumbers. The shift affects the peaks of the kinetic
energy (Fig.~\ref{fig:enspecs}) and the kinematic transfer functions
(Fig.~\ref{fig:ubcut}). Since the $e$-folding time of the magnetic
energy in the kinematic regime is significantly shorter than the
turnover time of the granular convection, most of the magnetic field
appears in the downflow lanes.

The shell-to-shell energy transfer functions (Fig.~\ref{fig:ubbou})
are similar in the kinematic state, in particular transfer from the
inertial range wavenumbers is dominant. This suggests that the dynamo
mechanism, namely the turbulent shear stress of the motions in the
inertial range, is essentially the same in each of the cases studied.
Neither the short correlation time of random forcing nor the
additional physics, as present in the Sun, alter the dynamo mechanism
previously discovered in incompressible turbulence with a mean flow.
As a property of the inertial range, independent of system-dependent
outer-scale circumstances, the dynamo mechanism is universal in the
Kolmogorov sense.

In the saturated state, the transfers involve much stronger energy
transfer from flows at the driving wavenumber. This is particularly
true for the strongly supercritical HoT and BC cases. Since the flow
at these wavenumbers is determined by the large-scale driving, the
dynamo action is no longer universal in the sense defined above.

The structure of magnetic fields produced by the small-scale dynamo
for magnetic Prandtl number $\Pram\ll1$ is different than for
$\Pram>1$ \citep{Schekochihin2007}.  We have shown that buoyancy and
stratification do not significantly alter the small-scale dynamo
mechanism.  Since small-scale dynamo action is known to work for
$\Pram\ll1$ in incompressible MHD \citep{Ponty2005,Schekochihin2007}
and with the same shell-to-shell transfer mechanism down to at least
$\Pram\approx0.5$ \citep{Alexakis2007a}, we expect small-scale dynamo
action to be possible for $\Pram\ll1$ both for \MURaM and the Sun.

\section{CONCLUSIONS}
\label{sec:conclusions}

We have compared simulations of solar mag\-ne\-to-convection and local
dynamo action with the properties of idealized systems in order to
evaluate the robustness and the range of applicability of the
conclusions drawn from studying the idealized models. The dynamo
has similar shell-to-shell energy transfer properties for
homogeneous-isotropic-incompressible turbulence, Boussinesq
convection, and solar conditions that include stratification,
compressibility, partial ionization and radiative energy transport.
The results suggest that the dynamo mechanisms, namely,
turbulent shear stresses acting in the inertial range, operate in the
same way in each of the cases considered.

For incompressible turbulence we find many similarities in the dynamo
generated by random forcing with a correlation time shorter than its
turnover time and that resulting from mean-flow driving previously
reported \citep{Alexakis:etal:2005,2005Mininni}.  While the
signature of the dynamo mechanism is the same, the role of
forcing-wavenumber fluid motions is diminished when
random forcing is used for the kinematic phase.  For the saturated
state, injection from forcing-wavenumbers remains significant even for
random forcing (similar to what was found by \citealt{2006Carati} for
yet another type of forcing).  Basic properties of the turbulent
small-scale dynamo process have been thoroughly studied for
homogeneous, isotropic, triply-periodic simulations; these properties
carry over to two situations that include more complex physics:
Boussinesq convection and solar surface convection.

\begin{acknowledgements}
\noindent This work has been supported by the Max-Planck Society in the
framework of the Interinstitutional Research Initiative \textit{Turbulent
transport and ion heating, reconnection and electron acceleration in
solar and fusion plasmas} of the MPI for Solar System Research,
Katlenburg-Lindau, and the Institute for Plasma Physics, Garching
(project MIF-IF-A-AERO8047).
\end{acknowledgements}

\bibliography{ref}

\begin{thebibliography}{35}
\expandafter\ifx\csname natexlab\endcsname\relax\def\natexlab#1{#1}\fi

\bibitem[{{Alexakis} {et~al.}(2005){Alexakis}, {Mininni}, \&
  {Pouquet}}]{Alexakis:etal:2005}
{Alexakis}, A., {Mininni}, P.~D., \& {Pouquet}, A. 2005, \pre, 72, 046301

\bibitem[{Alexakis {et~al.}(2007)Alexakis, Mininni, \& Pouquet}]{Alexakis2007a}
Alexakis, a., Mininni, P.~D., \& Pouquet, a. 2007, New Journal of Physics, 9,
  298

\bibitem[{{Batchelor}(1953)}]{1953Batchelor}
{Batchelor}, G.~K. 1953, {The Theory of Homogeneous Turbulence} (Cambridge
  University Press)

\bibitem[{{Brun} {et~al.}(2004){Brun}, {Miesch}, \& {Toomre}}]{Brun:etal:2004}
{Brun}, A.~S., {Miesch}, M.~S., \& {Toomre}, J. 2004, \apj, 614, 1073

\bibitem[{{Calzavarini} {et~al.}(2006){Calzavarini}, {Doering}, {Gibbon},
  {Lohse}, {Tanabe}, \& {Toschi}}]{Calzavarini:etal:2006}
{Calzavarini}, E., {Doering}, C.~R., {Gibbon}, J.~D., {Lohse}, D., {Tanabe},
  A., \& {Toschi}, F. 2006, \pre, 73, 035301

\bibitem[{{Carati} {et~al.}(2006){Carati}, {Debliquy}, {Knaepen}, {Teaca}, \&
  {Verma}}]{2006Carati}
{Carati}, D., {Debliquy}, O., {Knaepen}, B., {Teaca}, B., \& {Verma}, M. 2006,
  Journal of Turbulence, 7, 51

\bibitem[{{Cattaneo}(1999)}]{Cattaneo:1999}
{Cattaneo}, F. 1999, \apj, 515, L39

\bibitem[{{Cho}(2010)}]{2010Cho}
{Cho}, J. 2010, \apj, 725, 1786

\bibitem[{{Dar} {et~al.}(2001){Dar}, {Verma}, \& {Eswaran}}]{2001Dar}
{Dar}, G., {Verma}, M.~K., \& {Eswaran}, V. 2001, Physica D Nonlinear
  Phenomena, 157, 207

\bibitem[{{Debliquy} {et~al.}(2005){Debliquy}, {Verma}, \&
  {Carati}}]{2005Debliquy}
{Debliquy}, O., {Verma}, M.~K., \& {Carati}, D. 2005, Physics of Plasmas, 12,
  042309

\bibitem[{Eyink \& Aluie(2009)}]{Eyink2009}
Eyink, G.~L. \& Aluie, H. 2009, Physics of Fluids, 21, 115107

\bibitem[{{G{\'o}mez} {et~al.}(2005{\natexlab{a}}){G{\'o}mez}, {Mininni}, \&
  {Dmitruk}}]{Gomez:etal:2005a}
{G{\'o}mez}, D.~O., {Mininni}, P.~D., \& {Dmitruk}, P. 2005{\natexlab{a}},
  Advances in Space Research, 35, 899

\bibitem[{{G{\'o}mez} {et~al.}(2005{\natexlab{b}}){G{\'o}mez}, {Mininni}, \&
  {Dmitruk}}]{Gomez:etal:2005b}
---. 2005{\natexlab{b}}, Physica Scripta Volume T, 116, 123

\bibitem[{{Harris}(1978)}]{1978Harris}
{Harris}, F.~J. 1978, IEEE Proceedings, 66, 51

\bibitem[{{Kazantsev}(1968)}]{Kazantsev:1968}
{Kazantsev}, A.~P. 1968, Soviet Journal of Experimental and Theoretical
  Physics, 26, 1031

\bibitem[{{King} \& {Pringle}(2010)}]{King:Pringle:2010}
{King}, A.~R. \& {Pringle}, J.~E. 2010, \mnras, 404, 1903

\bibitem[{{Kleint} {et~al.}(2010){Kleint}, {Berdyugina}, {Shapiro}, \&
  {Bianda}}]{Kleint:etal:2010}
{Kleint}, L., {Berdyugina}, S.~V., {Shapiro}, A.~I., \& {Bianda}, M. 2010,
  \aap, 524, A37

\bibitem[{{Meneguzzi} {et~al.}(1981){Meneguzzi}, {Frisch}, \&
  {Pouquet}}]{Meneguzzi:etal:1981}
{Meneguzzi}, M., {Frisch}, U., \& {Pouquet}, A. 1981, Physical Review Letters,
  47, 1060

\bibitem[{{Mininni} {et~al.}(2005){Mininni}, {Alexakis}, \&
  {Pouquet}}]{2005Mininni}
{Mininni}, P., {Alexakis}, A., \& {Pouquet}, A. 2005, \pre, 72, 046302

\bibitem[{{Mininni} {et~al.}(2010){Mininni}, {Rosenberg}, {Reddy}, \&
  {Pouquet}}]{Mininni:etal:2010}
{Mininni}, P.~D., {Rosenberg}, D.~L., {Reddy}, R., \& {Pouquet}, A. 2010,
  arXiv:1003.4322v1 [physics.comp-ph]

\bibitem[{{Pietarila Graham} {et~al.}(2010){Pietarila Graham}, {Cameron}, \&
  {Sch{\"u}ssler}}]{Graham:etal:2010}
{Pietarila Graham}, J., {Cameron}, R., \& {Sch{\"u}ssler}, M. 2010, \apj, 714,
  1606

\bibitem[{{Pietarila Graham} {et~al.}(2009{\natexlab{a}}){Pietarila Graham},
  {Danilovic}, \& {Sch{\"u}ssler}}]{Graham:etal:2009b}
{Pietarila Graham}, J., {Danilovic}, S., \& {Sch{\"u}ssler}, M.
  2009{\natexlab{a}}, in Astronomical Society of the Pacific Conference Series,
  Vol. 415, Astronomical Society of the Pacific Conference Series, ed.
  {B.~Lites, M.~Cheung, T.~Magara, J.~Mariska, \& K.~Reeves}, 43

\bibitem[{{Pietarila Graham} {et~al.}(2009{\natexlab{b}}){Pietarila Graham},
  {Danilovic}, \& {Sch{\"u}ssler}}]{Graham:etal:2009a}
{Pietarila Graham}, J., {Danilovic}, S., \& {Sch{\"u}ssler}, M.
  2009{\natexlab{b}}, \apj, 693, 1728

\bibitem[{Ponty {et~al.}(2005)Ponty, Mininni, Montgomery, Pinton, Politano, \&
  Pouquet}]{Ponty2005}
Ponty, Y., Mininni, P., Montgomery, D., Pinton, J.-F., Politano, H., \&
  Pouquet, a. 2005, Physical Review Letters, 94, 2

\bibitem[{{Ryu} {et~al.}(2008){Ryu}, {Kang}, {Cho}, \& {Das}}]{Ryu:etal:2008}
{Ryu}, D., {Kang}, H., {Cho}, J., \& {Das}, S. 2008, Science, 320, 909

\bibitem[{{Schekochihin} \& {Cowley}(2006)}]{Schekochihin:Cowley:2006}
{Schekochihin}, A.~A. \& {Cowley}, S.~C. 2006, Physics of Plasmas, 13, 056501

\bibitem[{{Schekochihin} {et~al.}(2004){Schekochihin}, {Cowley}, {Taylor},
  {Maron}, \& {McWilliams}}]{Schekochihin:etal:2004}
{Schekochihin}, A.~A., {Cowley}, S.~C., {Taylor}, S.~F., {Maron}, J.~L., \&
  {McWilliams}, J.~C. 2004, \apj, 612, 276

\bibitem[{Schekochihin {et~al.}(2007)Schekochihin, Iskakov, Cowley, McWilliams,
  Proctor, \& Yousef}]{Schekochihin2007}
Schekochihin, a.~a., Iskakov, a.~B., Cowley, S.~C., McWilliams, J.~C., Proctor,
  M. R.~E., \& Yousef, T.~a. 2007, New Journal of Physics, 9, 300

\bibitem[{{Schleicher} {et~al.}(2010){Schleicher}, {Banerjee}, {Sur},
  {Arshakian}, {Klessen}, {Beck}, \& {Spaans}}]{Schleicher:etal:2010}
{Schleicher}, D.~R.~G., {Banerjee}, R., {Sur}, S., {Arshakian}, T.~G.,
  {Klessen}, R.~S., {Beck}, R., \& {Spaans}, M. 2010, \aap, 522, A115

\bibitem[{{Trujillo Bueno} {et~al.}(2004){Trujillo Bueno}, {Shchukina}, \&
  {Asensio Ramos}}]{Trujillo:etal:2004}
{Trujillo Bueno}, J., {Shchukina}, N., \& {Asensio Ramos}, A. 2004, \nat, 430,
  326

\bibitem[{{V{\"o}gler}(2003)}]{Voegler:2003}
{V{\"o}gler}, A. 2003, PhD thesis, University of G{\"o}ttingen, Germany,
  http://webdoc.sub.gwdg.de/diss/2004/voegler

\bibitem[{{V{\"o}gler} \& {Sch{\"u}ssler}(2007)}]{Voegler:Schuessler:2007}
{V{\"o}gler}, A. \& {Sch{\"u}ssler}, M. 2007, \aap, 465, L43

\bibitem[{{V{\"o}gler} {et~al.}(2005){V{\"o}gler}, {Shelyag}, {Sch{\"u}ssler},
  {Cattaneo}, {Emonet}, \& {Linde}}]{Voegler:etal:2005}
{V{\"o}gler}, A., {Shelyag}, S., {Sch{\"u}ssler}, M., {Cattaneo}, F., {Emonet},
  T., \& {Linde}, T. 2005, \aap, 429, 335

\bibitem[{{Waelkens} {et~al.}(2009){Waelkens}, {Schekochihin}, \&
  {En{\ss}lin}}]{Waelkens:etal:2009}
{Waelkens}, A.~H., {Schekochihin}, A.~A., \& {En{\ss}lin}, T.~A. 2009, \mnras,
  398, 1970

\bibitem[{{Wang} \& {Abel}(2009)}]{Wang:Abel:2009}
{Wang}, P. \& {Abel}, T. 2009, \apj, 696, 96

\end{thebibliography}

\appendix

\section{Shell-to-Shell Analysis for \MURaM}

The use of discrete Fourier transforms is complicated by the
non-periodicity of the \MURaM data in the vertical direction.  Ringing
effects may seriously taint the transfer functions. We prevent
ringing effects in this calculation by applying a 50\% Tukey
window \citep{1978Harris} on the data in the vertical direction before
the transfer analysis. In addition, we zero-pad beyond the extent of the
original data to exclude wrap-around effects.

We apply the usual shell filter decomposition \( \vec{a}(\vec{x}) =
\sum_K \vec{a}_K(x) \) to velocity $\vec{v}$, momentum $\rho\vec{v}$ and
the magnetic field $\vec{B}$, as described in Sect.~\ref{sec:transan}
but with a shell width of 4.  When computing the isotropic wavenumber
$K$, we relate all components of $\vec{k}$ to horizontal wavelengths,
i.e., \(K=1\) corresponds to one wave cycle in one of the two
axes-aligned horizontal directions.  The antisymmetry relations,
Eq.~\eqref{eq:anitcomp}, are satisfied analytically, both for periodic
boundary conditions and if a window is applied which tapers off to zero
at the boundary. In general, a surface integral contribution must be
applied.

\begin{figure*}[hb]
    \centering
    \includegraphics[width=.8\linewidth]{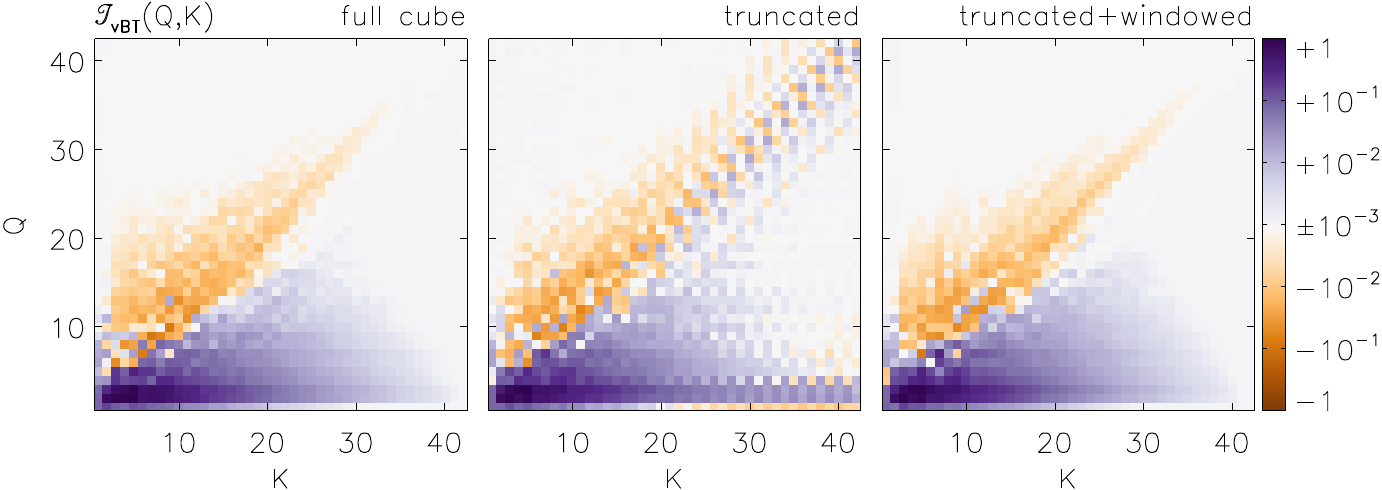}
\caption{Transfer function in a $128^3$ HoT simulation computed in three
different ways: using the original periodic cube (left), using a cube in
which half of the data has been zeroed (middle) and using a
``half-zeroed'' cube with the remaining data being forced to be
periodic by a Tukey window (right). The results have been renormalized
to account for the impact of the truncation and windowing on the
amplitude. }
\label{fig:wintest}
\end{figure*}

\begin{figure}[hb]
    \includegraphics[width=.5\linewidth]{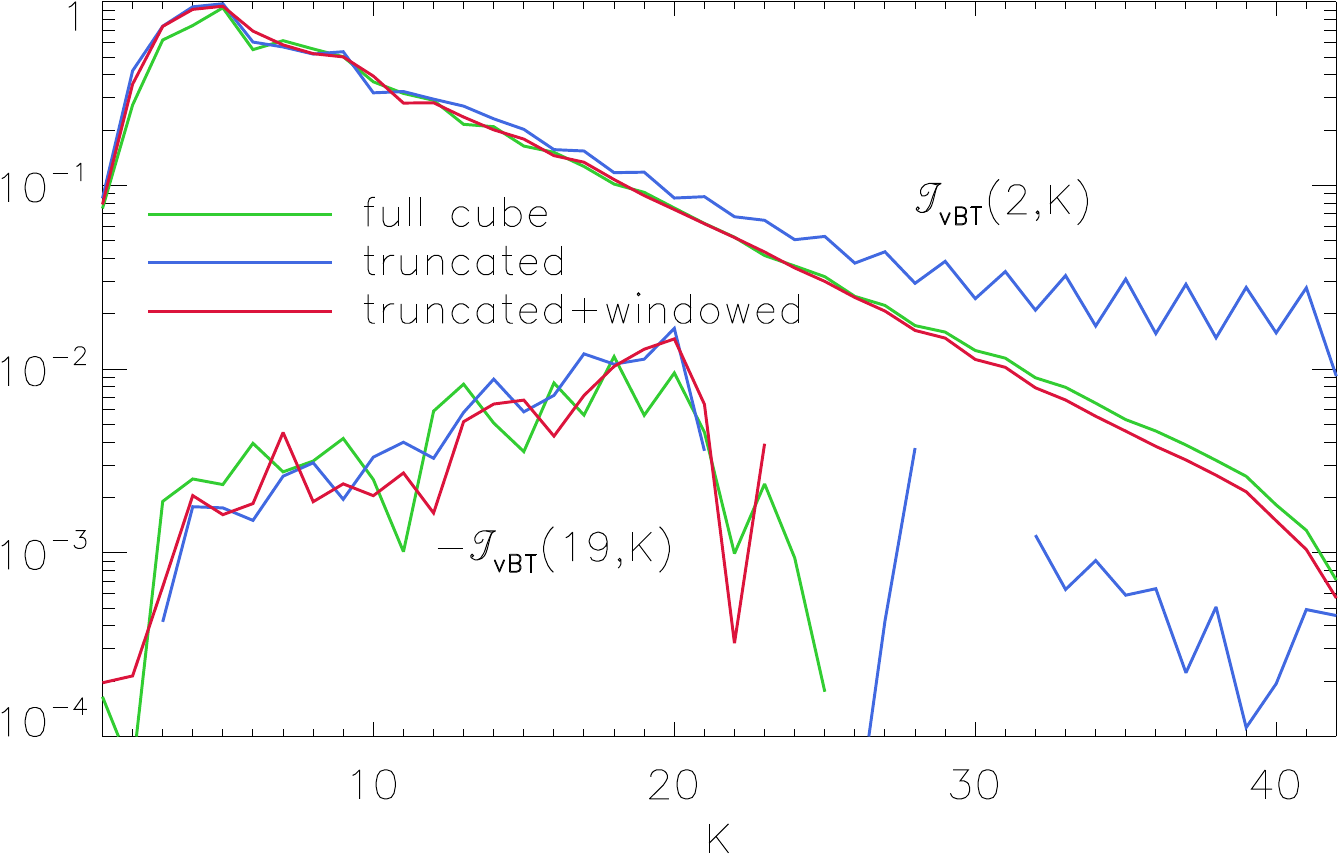}
\caption{Cuts through the transfer functions shown in Fig.~\ref{fig:wintest}.}
\label{fig:wintestcut}
\end{figure}

To assess the reliability of the transfers computed in this way, we
performed a series of tests using the results of a low-resolution HoT
run. In its original form, the data is fully periodic and the transfer
analysis is correct by definition.  If we ``truncate'' the data by
zeroing half of the box in one direction, the transfer analysis
results in spurious power at high frequencies in the form of
ringing noise, see Fig.~\ref{fig:wintest}.  If, in addition, a window is
applied to the data, the results obtained are  similar
to the original. As is illustrated in Fig.~\ref{fig:wintestcut}, the
slopes and amplitudes are reasonably well approximated.
The tests indicate that the transfer functions can be trusted within the
scope of the study presented in this paper.

The disparity between the vertical and horizontal extent of the
computational domain corrupts the transfer functions at low
wavenumbers \(K,Q \lesssim 7\). These large-scale modes
either do not exist in the vertical direction or are directly affected
by the alteration of the data by the window.  $K \approx 7$
corresponds to the vertical extent of the unaltered part of the data.
In the plots of the transfers for \MURaM (Figs.~\ref{fig:ubbou},
\ref{fig:ubcut}, and~\ref{fig:s2s_mur}), only wavenumbers above 8 are
shown.

\end{document}